\documentclass[12pt]{iopart}
\usepackage{epsfig}
\usepackage{iopams}
\newcommand{\Eins}
           {\;\smash{\raisebox{-0.5ex}{$\!\!\stackrel{\!\mbox{1}
            \hspace{-0.4ex}\rule[0.0ex]{0.06ex}{1.60ex}}{ }$}}}
\newtheorem{lemma}{Lemma}
\newtheorem{defi}{Definition}
\newtheorem{prop}{Proposition}
\newtheorem{ass}{Assumption}
\newtheorem{theorem}{Theorem}

\begin{document}

\title[Classical ground states]{Classical ground states of
symmetric Heisenberg spin systems}

\author{Heinz-J\"urgen Schmidt\dag
\footnote[3]{To
whom correspondence should be addressed (hschmidt@uos.de)}
and Marshall Luban\ddag}

\address{\dag\ Universit\"at Osnabr\"uck, Fachbereich Physik,
Barbarastr. 7, 49069 Osnabr\"uck, Germany}

\address{\ddag\ Ames Laboratory and Department of Physics and Astronomy, Iowa State University\\
Ames, Iowa 50011, USA}

\begin{abstract}
We investigate the ground states of classical Heisenberg spin systems which have
point group symmetry. Examples are the regular polygons (spin rings) and the seven quasi-regular polyhedra
including the five Platonic solids.
For these examples, ground states with special properties, e.~g.~coplanarity or symmetry,
can be completely enumerated using group-theoretical methods.
For systems having coplanar (anti-) ground states with vanishing total spin we also
calculate the smallest and largest energies of all states having a given total spin $S$.
We find that these extremal energies depend quadratically on S and prove that,
under certain assumptions, this happens
only for systems with coplanar $S=0$ ground states.
For general systems the corresponding parabolas
represent lower and upper bounds for the energy values. This provides
strong support and clarifies the conditions for the so-called
rotational band structure hypothesis which has
been numerically established for many quantum spin systems.
\end{abstract}



\maketitle

\section{Introduction}
The study of small interacting spin systems is not only of theoretical interest
but also of importance for the experimental investigation of recently synthetized
magnetic molecules \cite{G},\cite{TDPFGL},\cite{MPPG},\cite{MSSBSSKHTS}.
An exact calculation of the thermal expectation values of the
relevant quantum  observables and other quantities such as correlation functions is
possible only in very few cases.
Given this situation, a classical treatment yields a first approximation
for individual spins $s$ with $s\gg 1$ which is often astonishingly good,
as well as bounds for the exact quantum values (c.~f.~Berezin/Lieb-inequalities).
It is also valuable as a guide
for developing approximation schemes for attacking
the quantum theoretical problem.\\

One of the fundamental characteristics of a system is its ground state(s) and the
corresponding ground state energy.
The problem of determining the exact classical ground state
has been considered for a long time, see e.~g.~\cite{Ber},\cite{LK},
but has only been solved for a few special systems. These include arrays
of interacting spins occupying the sites of a regular $N$-polygon (spin rings)
with nearest-neighbor coupling constant $J>0$ (antiferromagnetic in our convention).
The corresponding ground state energy is
\begin{equation} \label{1}
E_{min}=-2 J N s^2, \mbox{    (even }N),
\end{equation}
corresponding to an alternating configuration of spin orientations in the $N$-polygon.
The analogous result
\begin{equation}\label{2}
E_{{\small min}}=2 J N s^2 \cos((N-1)\pi/N), \mbox{     (odd }N)
\end{equation}
does not appear to be in the literature, although
it is easily established, see (\ref{21}).
Another class of systems where the classical ground states are easily determined
are the so-called
bi-partite systems, which can be divided into two subsets $A$ and $B$ such that there
are only interactions (with a positive coupling constant $J$) between spins
$a\in A$ and $b\in B$. In this case any anti-parallel configuration between $A$ and
$B$ will minimize the energy. The regular polygons with even $N$ considered above and
the cube are simple examples of bi-partite systems.\\

Recently, the exact classical ground state of a system consisting
of $N=30$ spins occupying the vertices of an icosidodecahedron \cite{Cox},\cite{MW}
has been determined
\cite{AL}. The method employed there can be applied to any system which can be
decomposed into a set of triangles without common edges and which is 3-colorable.
Examples are
the octahedron (as mentioned in \cite{AL}) and the cuboctahedron (this is obtained
by joining the mid-points of the edges of a cube with their nearest neighbours, see
figure \ref{F-5}).
All these systems have coplanar ground states with $S=0$ which can be modified to remain
ground states in the presence of a magnetic field ${\cal H}$,
the corresponding energy depending quadratically on ${\cal H}$, c.~f.~\cite{AL}.
This yields a definite prediction for the magnetisation versus magnetic field
plot at low temperatures: It should have a constant slope until a critical magnetic
field ${\cal H}_c$ is applied, and for ${\cal H} > {\cal H}_c$ the magnetisation is saturated.\\

In the present paper we will derive a further consequence for systems having coplanar
ground states with $S=0$.
The same formal equation as that characterizing the ground state in the
presence of a magnetic field occurs for the ground state subject to the extra
constraint that the square of the total spin assumes a given value. Generalizing
the solution in \cite{AL} we obtain a whole family of states assuming the minimal
energy $E_{{\small min}}(S)$. It turns out that for these systems the function
$E_{{\small min}}(S)$
is quadratic in $S$ of the form
\begin{equation} \label{3}
E_{{\small min}}(S) =\frac{j-j_{{\small min}}}{N} S^2 +j_{{\small min}} N.
\end{equation}
Here $j$ is the row sum ($j=\sum_\nu J_{\mu \nu}$)
of the adjacency matrix ${\Bbb J}$ defined below. ${\Bbb J}$ is chosen
in such a way that $j$ does not
depend on $\mu$, and $j_{{\small min}}$ is the smallest eigenvalue of $ {\Bbb J}$.
Conversely, the existence of a lower parabola (\ref{3}) implies the
coplanarity of the ground state, see theorem \ref{CT1} for details.
Analogous results hold
for the maximal eigenvalue $E_{\small max}(S)$ of systems
with $S=0$ coplanar anti-ground states. Hence these systems show an exact rotational
band structure (RBS) which has been conjectured to apply for a general class of quantum spin
systems \cite{JM}. Thus our findings strongly support that conjecture in the sense
that the considered class of spin systems has exact RBS-parabolas in the classical
limit $s\rightarrow \infty$.
On the other hand we have found classes of systems which deviate from the exact
RBS-parabolas, especially for small values of $S$, see section 6.
Hence our work may be viewed as a first step towards an understanding of the conditions
for the occurence of RBS-parabolas.
We will call systems,
which exactly attain their lower RBS-parabola (\ref{3}), ``parabolic". For general systems
the corresponding parabolas represent lower and upper bounds for the energy values,
as will be proved in section 4. Sometimes these bounding parabolas will be very good
approximations of the boundaries of the energy spectrum even for non-parabolic systems.\\

Other questions arise in connection with spin systems having coplanar $S=0$
ground states.
Has every classical spin system coplanar
ground states or even coplanar $S=0$ ground states?
If not, what is the criterion for having coplanar ground states?
Are there systems with both coplanar and non-coplanar ground states?
We do not have complete answers to these questions,
but we can show the following:
\begin{itemize}
\item
In general the spin triangle with arbitrary coupling constants $J_1, J_2, J_3$
has only coplanar ground states with $S>0$.
\item
A spin system which we call ``pentagonal star" has a non-coplanar
ground state with $S>0$. The existence of coplanar states with the same
energy can be excluded on the basis of numerical calculations.
\item
For certain systems the existence of coplanar,
symmetric ground states can be excluded.  By ``symmetric" we mean roughly that the
state has the same symmetry as the spin system itself (for details see section 5).
\item Other systems, such as the dodecahedron and the icosahedron, have symmetric
non-coplanar ground states and apparently no coplanar ones. These ground states can be
geometrically vizualized in terms of
what is known \cite{Cox} as the ``great stellated dodecahedron"
and the ``great icosahedron", respectively.
\item The tetrahedron has a symmetric non-coplanar ground state with $S=0$ and a variety of
other non-symmetric ground states including coplanar ones, all of the same energy.
\item The same is true for other systems with full permutational symmetry (``$N$-pantahedron"),
except that these systems do not have symmetric ground states for $N>4$.
\item The cuboctahedron has a lower RBS-parabola but interestingly it
has two symmetric ground states with $S=0$, a coplanar one
and a non-coplanar one. The latter corresponds to a degenerate stellated
figure which we have not found in the literature
and which is composed of four Stars of David, see figure \ref{F-6}.
\end{itemize}

Thus we are faced with a surprising variety of possibilities of ground states of different
types. The main open problem is to rigorously prove the existence of
spin systems without coplanar ground states.
It appears that quantitative statements can be made mainly if symmetric states are concerned,
since in this case we may use standard methods of group theory, in particular representation
theory of the point groups. \\

The article is organized as follows: Section 2 contains the general definitions and results
which are independent of symmetry assumptions. The main result is that systems with coplanar
$S=0$ ground states have RBS-parabolas (theorem 1) and its converse (theorem 2).
These parabolas represent lower and upper energy
bounds even for non-parabolic systems, as proven in section 4. In section 3
we address questions of extension of ground states to larger systems and ``higher order frustration".
Section 5 introduces symmetry assumptions
and the definition and simple properties of symmetric states. This machinery is applied in section 6 to
the investigation of particular examples.
We consider the general triangle which is generically non-parabolic
as well as classes of symmetric spin systems including the quasi-regular polyhedra. We construct symmetric
ground states from certain irreducible representations of the respective point groups.
The article closes with a summary.

\section{Notations, definitions, and general results}
We consider as the phase space of the classical spin system the $N$-fold product
of the unit sphere ${\cal S}^2$. Instead of using canonical coordinates we will denote a state
by a sequence $\bi{s}$ of unit vectors $\bi{s}_\mu, \mu=1,\ldots,N$ with components
 ${s}_\mu^{i}, i=1,2,3$.
The total spin is $\bi{S}\equiv \sum_\mu  \bi{s}_\mu$ with components
$\bi{S}^{i}, i=1,2,3$.
If not indicated otherwise, the abstract letters $i,j,\ldots$ will always be
(upper) indices and there is no danger of confusion with the square of a vector,
e.~g.~$\bi{S}^2$. We will use bracketed indices to denote by, say,
$x^{(i)}$ the vector with the components $x^{i}$, similarly for matrices. For
example, the $n\times m$ matrix $A_{(i)(j)}$ consists of $n$ row vectors
$A_{i(j)}$ and of $m$ column vectors $A_{(i)j}$. \\

If not mentioned otherwise, the Hamilton function (or ``energy") will be of the form
\begin{equation}  \label{4}
H_0(\bi{s})=\sum_{\mu \nu} J_{\mu\nu} \bi{s}_\mu \cdot \bi{s}_\nu.
\end{equation}
If a magnetic field $\vec{\cal H}$ is to be included we add a
Zeeman term and obtain
\begin{equation}\label{4a}
H_h=H_0 - \bi{h}\cdot \bi{S},
\end{equation}
where $\bi{h}\equiv \sqrt{s(s+1)} g\mu_B \vec{\cal H}$ contains the common combination of the
spectroscopic splitting factor $g$ and the Bohr magneton $\mu_B$ and is scaled with the
function $\sqrt{s(s+1)}$ of the corresponding spin quantum number $s$.

Note that the exchange parameters $J_{\mu\nu}$ are not uniquely determined
by the Hamiltonian $H_0$ via (\ref{4}). Different choices of the $J_{\mu\nu}$
leading to the same $H_0$ will be referred to as different ``gauges".
We will adopt the following gauges.
First, the antisymmetric part of ${\Bbb J}$ does not occur in the Hamiltonian
(\ref{4}). Hence we will follow common practice and
choose $J_{\mu\nu}=J_{\nu\mu}$. Thus
the $J_{\mu\nu}$ can be considered as the entries of a real symmetric matrix $\Bbb J$. Second,
since  ${\bf s}_\mu\cdot{\bf s}_\mu =1$ we may choose arbitrary
diagonal elements  $J_{\mu\mu}$ without changing $H_0$, as long as their sum
vanishes, $\mbox{Tr}{\Bbb J}=0$. The usual gauge chosen throughout the
literature is  $J_{\mu\mu}=0,\; \mu=1,\ldots,N,$ which will be called the
``zero gauge". In this article, however, we will choose another gauge,
to be called the ``homogeneous gauge", which is defined by the condition that the row sums
\begin{equation}\label{4b}
J_\mu \equiv \sum_\nu  J_{\mu\nu}
\end{equation}
will be independent of $\mu$. Note that the eigenvalues of ${\Bbb J}$
may non--trivially depend on the gauge.
We found that the eigenvalues of ${\Bbb J}$ are only relevant for energy estimates
if the row sum of ${\Bbb J}$ is constant. This would restrict the applicability of
large parts of our theory if we adopt the zero gauge. However, by introducing the
homogeneous gauge we can apply our results to a very general class of systems.\\

The quantity
\begin{equation}\label{4c}
Nj\equiv\sum_{\mu\nu}  J_{\mu\nu}
\end{equation}
is gauge--independent. If exchange parameters satisfying
$\widetilde{J}_{\mu\nu}=\widetilde{J}_{\nu\mu}$
are given in the zero gauge, the corresponding
parameters $J_{\mu\nu}$ in the homogeneous gauge are obtained as follows:
\begin{equation}\label{4d}
J_{\mu\nu} \equiv \widetilde{J}_{\mu\nu} \mbox{ for } \mu \neq \nu,
\end{equation}
\begin{equation}\label{4e}
J_{\mu\mu} \equiv j - \widetilde{J}_\mu.
\end{equation}
It follows that
\begin{equation}\label{4f}
j=J_\mu =\sum_\nu \widetilde{J}_{\mu\nu} + J_{\mu\mu} .
\end{equation}

A spin system is called \underline{antiferromagnetic} (AF) iff all
$J_{\mu\nu}\ge 0$ for $\mu\neq\nu$. A system is called \underline{connected}
iff its spin sites cannot be decomposed into two disjoint classes, say ${\cal A}$
and ${\cal B}$, such that $J_{\mu\nu}=0$ if $\mu\in{\cal A}$ and $\nu\in{\cal B}$.\\

In the special case where all exchange parameters
$J_{\mu\nu},\; \mu\neq\nu$ are $0$ or $J$, the system
can be essentially represented by its graph $\Gamma$.
A graph $\Gamma=({\cal {V,E}})$ consists of a set of ``vertices" ${\cal V}$
and a set of ``edges" ${\cal E}\subset \left[{\cal V}\right]^2$. Here
$\left[{\cal V}\right]^2$ denotes the set of subsets of ${\cal V}$
with exactly $2$ elements.
In our case,
the vertices are the spin sites, ${\cal V}=\{1,\ldots,N\}$, and the edges
represent interacting pairs of spins.
For spin systems it is appropriate to consider only
graphs without loops and multiple edges, as we do in this article,
following, e.~g.~, \cite{Die}. \\

There are different ways to graphically represent a spin state ${\bi s}_\mu$.
For specific spin systems, e.~g.~magnetic molecules, the spin sites
$\mu\in{\cal V}$ are embedded into the physical 3-space and represented by
vectors ${\bi r}_\mu$. One way would be to attach the vectors ${\bi s}_\mu$
as small arrows to the sites given by ${\bi r}_\mu$. This is done in figure
\ref{F-8} below. However, when using this representation it is difficult to
visualize the structure of the ${\bi s}_\mu$ in spin space. Hence we will mostly
employ another method of representation: We draw the $N$ vertices in
spin space according to the unit vectors ${\bi s}_\mu$
and join them with lines according to the edges of the original spin
graph $\Gamma$. Thus we obtain a graph isomorphic to $\Gamma$ which contains
the additional information about the considered spin configuration.
\\

In graph theory the set of edges is often represented by a matrix called the
\underline{adjacency matrix}. We will use this name also for the matrix ${\Bbb J}$
in the general case of different exchange parameters. There exists an extended
literature about the connection between the structure of a graph and the spectrum of
${\Bbb J}$, see, for example \cite{CDGT}, where also applications in chemistry
and physics are mentioned.\\

The graph $\Gamma$ is called \underline{complete} if
${\cal E}=\left[{\cal V}\right]^2$. We will also call the corresponding spin system
where any two spins interact with equal strength, a \underline{pantahedron}.\\

Being symmetrical, ${\Bbb J}$ has a complete set of $N$ real, ordered eigenvalues
$j_{\small min},\ldots,j_{\small max}$ which are counted according to their multiplicity.
One of them is the row sum $j$ with
$\bi{1}\equiv\frac{1}{\sqrt{N}}(1,1,\ldots,1)$ as the corresponding
eigenvector.
For connected AF-systems, $j$ will be non-degenerate and equals the largest
eigenvalue, $j=j_{\small max}>0$, by the theorem of Frobenius-Perron.
It will be convenient to rearrange the indices such that $j=j_0$ and the
remaining eigenvalues $j_1,\ldots,j_{N-1}$ are ordered according to their size.
Note that for AF systems, $j_1=j_{\small min}$ since $j_{\small min}=j>0$ contradicts
$\Tr {\Bbb J}=0$. However, since it may happen that $j_{\small max}=j$ we will
write $j_{\small maxi}$ for the largest eigenvalue of ${\Bbb J}$ other than $j$,
counted once, hence
$j_{N-1}=j_{\small maxi}$ for AF systems.
Sums over $\alpha=1,\ldots,N-1$ but excluding
$\alpha=0$ will be denoted by $\sum'$.
We will denote
the $\alpha$-th normalized eigenvector of ${\mathbb J}$ by
$C^{(\nu)}_{\alpha}$, i.~e.~
\begin{equation}\label{B1}
\sum_{\nu}J_{\mu\nu} C^\nu_{\alpha} =
j_\alpha C^\mu_{\alpha},\quad \mu,\alpha=0,\ldots,N-1,
\end{equation}
and
\begin{equation}\label{B2}
\sum_\mu \overline{C^\mu_{\alpha}} C^\mu_{\beta}=
\delta_{\alpha\beta},  \quad \alpha,\beta=0,\ldots,N-1,
\end{equation}
where we also allow for the possibility to choose complex
eigenvectors. \\

We note in passing that ${\Bbb J}$ in the homogeneous gauge
is essentially the Hamilton operator of the
corresponding quantum spin system, restricted to the subspace of total magnetic
quantum number $M=N s -1$. Thus there is an unexpected connection between ``weakly
symmetric" classical states to be defined below and quantum eigenstates of the Hamiltonian
lying in the mentioned subspace.\\

The equations of motion resulting  from the Hamiltonian (\ref{4}) are
\begin{equation}\label{5}
\frac{d}{dt}\bi{s}_\mu=\left(\sum_\nu J_{\mu\nu}\bi{s}_\nu\right)\times  \bi{s}_\mu.
\end{equation}
Using $J_{\mu\nu}=J_{\nu\mu}$ one can immediately show that
the total spin vector $\bi{S}$ is a constant of motion.\\

We now will define various kinds of special states. A \underline{stationary}
state will be one with $\frac{d}{dt}\bi{s}_\mu=0, \mbox{  for all  }
\mu=1,\ldots,N$. According to (\ref{5}), this is equivalent to
\begin{equation}\label{6}
\sum_\nu J_{\mu\nu}\bi{s}_\nu\ = \kappa_\mu \bi{s}_\mu,
\end{equation}
for some real numbers $\kappa_\mu,\mu=1,\ldots,N$.
Equation (\ref{6}) can also be viewed as the solution of the problem
to seek states with vanishing  variation of the quadratic form
\begin{equation}\label{7}
\sum_{\mu\nu} J_{\mu\nu}\bi{s}_\mu\cdot\bi{s}_\nu
\end{equation}

\noindent
subject to the constraints
$\bi{s}_\mu\cdot\bi{s}_\mu=1,\quad\mu=1,\ldots,N$.  The $\kappa_\mu$ then appear
as Lagrange parameters of this variational problem.\\

Geometrically, (\ref{6}) means that
for a stationary state each spin vector is proportional to the ``center of mass"
of its neighbors. As for general Hamiltonian systems, the stationary states are
just the critical points of the Hamilton function, i.~e.~those points where the
gradient of $H_0$ vanishes. In particular, the states  $\bi{s}$
with minimal $H_0(\bi{s})=E_{{\small min}}$ or maximal $H_0(\bi{s})=E_{max}$ are stationary.
The former will be called
\underline{ground states}, and the latter \underline{anti-ground states}.
For AF systems the anti-ground state is always
of the form  $\bi{s}_\mu=\bi{e}$ for all $\mu=1,\ldots,N$, i.~e.~all spins are
aligned parallel to an arbitrary unit vector $\bi{e}$ (ferromagnetic ordering).
In this case $j_{max}=j$  and $E_{max}=N j$.\\

Moreover, we will consider the \underline{relative ground states},
(\underline{relative anti-ground states}), which are defined  as the states
of minimal (maximal) energy among all states satisfying
$\left(\sum_\nu \bi{s}_\nu\right)^2=S^2$. Let $E_{{\small min}}(S)$ denote the energies of the
relative ground states, and  $E_{max}(S)$ that of the relative anti-ground states. Again,
the relative ground states and anti-ground states are among the
solutions of the variational problem
corresponding to (\ref{7}) with the additional constraint
$\left(\sum_\nu \bi{s}_\nu\right)^2=S^2>0$. If the extra Lagrange parameter is called
$\chi$, we obtain the condition, analogous to (\ref{6}),
\begin{equation} \label{8}
\sum_\nu J_{\mu\nu}\bi{s}_\nu\ = \kappa_\mu \bi{s}_\mu + \chi \bi{S}\quad \mbox{ for } S>0.
\end{equation}
The equations of motion (\ref{5}) then imply that for states satisfying (\ref{8})
we have
\begin{equation}\label{9}
\frac{d}{dt}\bi{s}_\mu=\chi \bi{S}\times  \bi{s}_\mu,
\end{equation}
i.~e.~, for $S>0$ these states are no longer stationary but are
precessing around the total spin vector.
This is completely analogous to the precession of stationary states of $H_0$
in the presence of a magnetic field.\\

The condition $S>0$ in (\ref{8}) is necessary in order to apply the method of Lagrange
parameters. $S=0$ has to be excluded since it would correspond to a critical point
of the constraining function $g({\bi s})= (\sum_\nu {\bi s}_\nu)^2 -S^2$, see for example
\cite{AMR}. To cover also the case $S=0$ we consider three constraining equations written
in vector form
\begin{equation}\label{9a}
\sum_\nu {\bi s}_\nu  = {\bi S}_0,
\end{equation}
where ${\bi S}_0$ is a fixed vector. Then the above-mentioned problem does not occur.
We have three Lagrange parameters which will be written as the components of a vector ${\bi L}$.
The resulting equations are
\begin{equation}\label{9b}
\sum_\nu J_{\mu\nu}{\bi s}_\nu  = \kappa_\mu {\bi s}_\mu + {\bi L}.
\end{equation}
Taking the vector product of (\ref{9b}) $\times {\bi s}_\mu$ and summing over $\mu$
yields ${\bi 0}= {\bi S}\times {\bi L}$. If ${\bi S}\neq {\bi 0}$ this is equivalent
to ${\bi L}= \chi {\bi S}$ and we recover (\ref{8}). For $S=0$, however, (\ref{8})
has to be replaced by (\ref{9b}).\\

States satisfying (\ref{9b})
will be called \underline{weakly stationary} states, since they are stationary
in a rotating frame.
Hence relative (anti-) ground states are weakly stationary.\\

There exists a simple mechanical model for stationary states: Consider a system of $N$
rigid, massless rods of unit length fixed at the same point (``pendula").
These rods may be represented by unit vectors ${\bi s}_\mu$. Between two rods
${\bi s}_\mu$ and ${\bi s}_\nu$ we consider springs satisfying Hooke's law with
spring constants $2 J_{\mu\nu}$. Further there is a constant (``gravitational") force
${\bi F}= F {\bi e}$. By the cosine theorem,
$|{\bi s}_\mu - {\bi s}_\nu|^2 = 2(1- {\bi s}_\mu \cdot {\bi s}_\nu )$, hence
the potential energy of our mechanical model will be
\begin{equation}\label{9c}
V({\bi s}) =\frac{1}{2} \sum_{\mu\nu} J_{\mu\nu}
|{\bi s}_\mu - {\bi s}_\nu|^2 -{\bi F}\cdot {\bi S}
=
Nj-\sum_{\mu\nu} J_{\mu\nu}  {\bi s}_\mu \cdot {\bi s}_\nu
 -{\bi F}\cdot {\bi S}.
\end{equation}
The equilibrium points of this mechanical system will satisfy the zero-force
equation
\begin{equation}\label{9d}
2\sum_{\mu\nu} J_{\mu\nu} {\bi s}_\nu + {\bi F} = 2\kappa_\mu {\bi s}_\mu,
\end{equation}
where $2\kappa_\mu {\bi s}_\mu$ are essentially
the constraining forces exerted on the rigid rods.
Moreover, for equilibrium ${\bi S}$ will be proportional to the constant force,
say ${\bi F}=-2 \chi {\bi S}$, since otherwise there
would be a first-order variation of the potential energy when performing infinitesimal
uniform rotations of the system. It follows that the equations (\ref{9d}) and
(\ref{8}) are formally identical. Summarizing, there is a $1:1$ correspondence between
the weakly stationary spin states ${\bi s}$ and the equilibrium states  ${\bi s}$
of the considered mechanical systems, if the value of the force $F$ runs through
all real numbers. This mechanical model may help the reader
to vizualize some of the weakly stationary
states to be considered in this article.\\


We can easily obtain the following bounds for the energy of stationary states $\bi{s}$:
\begin{lemma}\label{L1}
$N j_{{\small min}}  \le E_{{\small min}}\le H_0(\bi{s})=\sum_\mu \kappa_\mu
\le E_{\small max} \le N j_{\small max}. $   Moreover, for AF systems and $S=0$ we have
$H_0({\bi s})\le N j_{\small maxi}$.
\end{lemma}
{\bf Proof:} $H_0(\bi{s})=\sum_{\mu\nu}J_{\mu\nu}\bi{s}_\nu\cdot\bi{s}_\mu=
\sum_{\mu}\kappa_\mu \bi{s}_\mu\cdot\bi{s}_\mu= \sum_\mu \kappa_\mu.$ Further we have
$\sum_{\mu\nu}J_{\mu\nu}\bi{s}_\nu\cdot\bi{s}_\mu \le
j_{max}  \sum_{\mu}\bi{s}_\mu\cdot\bi{s}_\mu = N j_{max}$, and analogously for the
lower bound. Since (anti-) ground states are stationary,
these bounds also hold for $E_{{\small min}}$ and $E_{max}$. The upper energy bound
for $S=0$ and AF systems follows from
$\sum_{\mu\nu}J_{\mu\nu}\bi{s}_\nu\cdot\bi{s}_\mu \le
j_{\small maxi}  \sum_{\mu}\bi{s}_\mu\cdot\bi{s}_\mu$, since for ${\bi S}={\bi 0}$
the vectors $s_{(\mu)}^i,\; i=1,2,3$ are orthogonal to the eigenvector ${\bi 1}$
corresponding to the eigenvalue $j$.
\hspace*{\fill}\rule{3mm}{3mm}  \\

The respective bounds of lemma 1 are attained if there exist stationary states where all
$\kappa_\mu=j_{{\small min}}$ or $\kappa_\mu=j_{\small max}$, or
 $\kappa_\mu=j_{\small maxi}$ in the AF case for $S=0$. At present, we cannot show the
existence of such states under general conditions but this
can be done for some specific examples.
We will call any stationary state with the property that the values of all $\kappa_\mu$
of (\ref{6}) are independent of $\mu$, a \underline{weakly symmetric}
state. If we view the set of components $s_\mu^{i}$ of a state $\bi{s}$
as an $N\times 3$-matrix, the definition of ``weakly symmetric" could be rephrased
in the way that all rows of $\bi{s}$ are eigenvectors of ${\Bbb J}$ for the same eigenvalue
$\widetilde{j}$. Recall that for stationary states each spin vector  $\bi{s}_\mu$
is proportional to the ``center of mass" of the neighbouring spins.
For weakly symmetric states the constant
of proportionality will be the same for all spins $\bi{s}_\mu$, which motivates the
wording. Obviously, weakly symmetric (anti-)ground states belong to the eigenvalue
$j_{{\small min}}$, ($j_{\small max}$). If one takes any three eigenvectors of ${\Bbb J}$
with the same eigenvalue to form the 3 rows of a matrix, one will not get
automatically a weakly symmetric state, since the $N$ columns need not be unit vectors.
In section 5 we will present a construction procedure which,
under certain assumptions, yields (weakly) symmetric states.\\

Since the energy $H_0({\bi s})$ is invariant w.~r.~t.~rotations/reflections in spin space the above-defined
classes of weakly symmetric states, (weakly) stationary states, or of (relative) ground states are also
invariant under rotations/reflections.\\

Another simple property of weakly symmetric ground states is the following:
\begin{lemma}\label{L2}
Every weakly symmetric state has vanishing total spin if the corresponding
eigenvalue $\widetilde{\j}$ is different from the row sum $j$. In particular,
every weakly symmetric ground state of an AF system has vanishing total spin.
\end{lemma}
{\bf Proof}: We obtain
$\sum_{\mu\nu}J_{\mu\nu} \bi{s}_\nu= j \sum_{\nu}\bi{s}_\nu = j \bi{S}$
and
$\sum_{\mu\nu}J_{\mu\nu} \bi{s}_\nu= \widetilde{\j} \sum_{\mu}\bi{s}_\mu = \widetilde{\j} \bi{S}$.
Hence $\widetilde{\j}=j$ or $\bi{S}=\bi{0}$.
If $\bi{s}$ is a weakly symmetric ground state of an AF system
with $\widetilde{\j}=j_{{\small min}}$, we have
$j_{{\small min}}<j$ as above.
\hspace*{\fill}\rule{3mm}{3mm}  \\

We will call a state $\bi{s}$ \underline{collinear}
if all its spin vectors are (anti-)parallel
to a given vector $\bi{e}$, i.~e.~$\bi{s}_\mu=\pm\bi{e}$.
Due to equation (\ref{6}), every collinear state is stationary.
Further we will call a state $\bi{s}$ \underline{coplanar}
if all its spin vectors ly in a plane
i.~e.~, if there exists a non-zero 3-vector $\bi{n}$ such that
$\bi{n}\cdot\bi{s}_\mu=0$ for all
$\mu=1,\ldots,N$. In section 6 we will show that systems
such as the regular polygon,
the pantahedron and the quasi-regular cuboctahedron and icosidodecahedron
admit coplanar weakly symmetric $S=0$ ground states. These ground states are important
since one can construct from them a whole family of relative ground states,
which show a rotational band structure.
\begin{theorem}\label{T1}
\begin{enumerate}
\item
Let $\bi{s}$ be a coplanar weakly symmetric state with $S=0$ corresponding to an eigenvalue
$\widetilde{\j}$ of ${\Bbb J}$ and perpendicular to a unit vector $\bi{n}$. Then the following
family of states
$\hat{\bi{s}}(S), 0\le S\le N$
has total spin length S:
\begin{equation}\label{10}
\hat{\bi{s}}_\mu(S)=\sqrt{1-\frac{S^2}{N^2}}\bi{s}_\mu+\frac{S}{N}\bi{n},
\quad \mu=1,\ldots,N.
\end{equation}
\item If, moreover, $\bi{s}$ is a ground state then $\hat{\bi{s}}(S)$ will be
a relative ground state for all $0\le S\le N$.
In this case,
\begin{equation}\label{11}
E_{{\small min}}(S)=\frac{j-j_{{\small min}}}{N}S^2+j_{{\small min}}N.
\end{equation}
\item  The analogous case for  $\bi{s}$ being a  coplanar weakly symmetric  relative
anti-ground state for $S=0$:
Then $\hat{\bi{s}}(S)$ will be
a relative anti-ground state and
\begin{equation}\label{12}
E_{max}(S)=\frac{j-j_{\small maxi}}{N}S^2+j_{\small maxi}N.
\end{equation}
\end{enumerate}
\end{theorem}
{\bf Proof}:
\begin{enumerate}
\item Obviously, $\hat{\bi{s}}(S)$ is a state with total spin length $S$.
\item We obtain
$\sum_{\nu}  J_{\mu\nu}    \hat{\bi{s}}_\nu =
\sqrt{1-\frac{S^2}{N^2}}j_{{\small min}}\bi{s}_\mu+\frac{S}{N}j\bi{n}.$
This is of the form (\ref{8}) if we choose $\chi=\frac{j-j_{{\small min}}}{N}$.
In order to prove that $\hat{\bi{s}}$ is a relative ground state, we note
that $\hat{\bi{s}}$ is a weakly symmetric ground state w.~r.~t.~the modified
adjacency matrix $\acute{J}_{\mu\nu}\equiv J_{\mu\nu}-\chi$, since $j_{{\small min}}$
is also the smallest eigenvalue of $\acute{{\Bbb J}}$. The calculation of $E_{{\small min}}(S)$
is straightforward.
\item This case is largely analogous, but $j_{{\small min}}$ has to be replaced by $j_{\small maxi}$,
since $\bi{s}$ has $S=0$ and cannot be the total anti-ground state.
\end{enumerate}
\hspace*{\fill}\rule{3mm}{3mm}  \\

Systems which satisfy equation (\ref{11})
will be called \underline{parabolic} systems. Hence the essential claim of
theorem \ref{T1} is that systems with a coplanar $S=0$ ground state will be parabolic.
\\

We will now prove a theorem which can be viewed as the converse of theorem 1, in so far
it shows that coplanar ground states necessarily appear for parabolic systems, if certain
technical assumptions are satisfied. We expect that these assumptions hold under fairly
general conditions but will not dwell upon this question further.

\begin{theorem}\label{CT1}
Consider a connected AF system and let $t\mapsto \mathbf{s}(t)$ be a smooth curve for $0<t<\epsilon$ of relative ground states
such that the limits $t\rightarrow 0$ of $\mathbf{s}(t)$
and its derivatives up to second order exist, in particular
$\frac{dS}{dt}(0)\neq 0$ for $S^2(t)\equiv(\sum_\mu \mathbf{s}_\mu(t))^2$.
Moreover, let $\mathbf{s}(0)$ be a weakly symmetric ground state with $S=0$ and assume parabolicity,
$E_{\small min}(S) = \frac{j-j_{\small min}}{N} S^2 +N j_{\small min}$, at least for
the interval covered by $0<t<\epsilon$.\\
Then  $\mathbf{s}(0)$ will be coplanar.
\end{theorem}
{\bf Proof}:
We will indicate the limit $t\rightarrow 0$ of a function of $t$
by skipping the argument, e.~g.~${\bi s}_\mu\equiv{\bi s}_\mu(0)$.
Differentiation with respect to $t$ will be indicated by an overdot.
Since ${\bi s}_\mu(t)\cdot{\bi s}_\mu(t)=1$, differentiation yields
\begin{equation} \label{C0}
{\bi s}_\mu\cdot\dot{\bi s}_\mu=0, \quad
\dot{\bi s}_\mu\cdot\dot{\bi s}_\mu+{\bi s}_\mu\cdot\ddot{\bi s}_\mu=0.
\end{equation}
\noindent
We may assume that ${\bi S}(t)=S(t) {\bi e}$ where ${\bi e }$ is a constant unit vector.
Being a weakly stationary relative ground state,  $\mathbf{s}(t)$  satisfies  (\ref{8}) for $t>0$:
\begin{equation} \label{C1}
{\bi j}_\mu(t)\equiv\sum_\nu J_{\mu\nu} {\bi s}_\nu(t) =
\kappa_\mu(t) {\bi s}_\mu(t) + \chi(t) \sum_\nu {\bi s}_\nu (t),
\end{equation}
with the limit value ${\bi S} \equiv \sum_\nu {\bi s}_\nu = {\bi 0}$
according to the assumptions of the theorem. But we cannot a priori assume that $\chi(t)$ has a finite
limit value for  $t\rightarrow 0$. Thus we solve (\ref{C1}) for $L(t)\equiv S(t) \chi(t)$ and obtain
\begin{equation} \label{C1a}
L(t)=\frac{{\bi j}_\mu(t)\cdot {\bi e} -{\bi j}_\mu(t)\cdot {\bi s}_\mu(t)\; {\bi e}\cdot {\bi s}_\mu(t)}
{1-({\bi e}\cdot {\bi s}_\mu(t))^2}.
\end{equation}
\noindent
If ${\bi s}$ is collinear, the proof is done. If ${\bi s}$ is not collinear, we find some $\mu$ such that
$({\bi e}\cdot {\bi s}_\mu(t))^2 \neq 1$ for all $0\le t <\epsilon$.
For this value of $\mu$, (\ref{C1a}) defines a smooth function
of $t$, independent of $\mu$,  with a finite limit value $L\equiv L(0)$.
Moreover, we can solve (\ref{C1}) for $\kappa_\mu(t)$, $\mu=1,\ldots,N$ and obtain
\begin{equation} \label{C1b}
\kappa_\mu(t)=   \sum_\nu J_{\mu\nu} {\bi s}_\nu(t)\cdot{\bi s}_\mu(t)  -
L(t) {\bi e}\cdot {\bi s}_\mu(t).
\end{equation}
\noindent
This shows that also  $\kappa_\mu(t)$ is a smooth function of $t$ and has a finite limit
for $t\rightarrow 0$.  Since, by assumption, ${\bi s}$ is weakly symmetric this limit must be
$\kappa_\mu=j_{\small min}$ and hence $L=0$. Therefore the smooth function
$\chi(t)=\frac{L(t)}{S(t)}$ for $t>0$ has a finite limit
$\chi=\lim_{t\rightarrow\infty} \frac{\dot{L}(t)}{\dot{S}(t)}$, employing  l'Hospital's rule
and that $\dot{S}(0)\neq 0$ by assumption.\\

Differentiating (\ref{C1}) twice and
taking the limit $t\rightarrow 0$ yields
\begin{equation} \label{C2}
\sum_\nu J_{\mu\nu} \dot{\bi s}_\nu =
\dot{\kappa}_\mu {\bi s}_\mu + j_{\small min} \dot{\bi s}_\mu + \chi \dot{\bi S},
\end{equation}
and
\begin{equation} \label{C3}
\sum_\nu J_{\mu\nu} \ddot{\bi s}_\nu =
\ddot{\kappa}_\mu {\bi s}_\mu +
2\dot{\kappa}_\mu \dot{\bi s}_\mu  +
j_{\small min} \ddot{\bi s}_\mu + 2\dot{\chi} \dot{\bi S}+{\chi} \ddot{\bi S}.
\end{equation}
Multipying (\ref{C1}),(\ref{C2}), and (\ref{C3})  with ${\bi s}_\mu$ and
summing over $\mu$ yields
\begin{equation} \label{C4}
H=\sum_{\mu\nu} J_{\mu\nu} {\bi s}_\mu\cdot{\bi s}_\nu =
\sum_\mu{\kappa}_\mu + \chi {\bi S}^2 = N j_{\small min},
\end{equation}
\begin{equation} \label{C5}
\frac{1}{2}\dot{H}=\sum_{\mu\nu} J_{\mu\nu} \dot{\bi s}_\nu\cdot{\bi s}_\mu =
\sum_\mu\dot{\kappa}_\mu =
\sum_{\nu} j_{\small min} {\bi s}_\nu\cdot\dot{\bi s}_\nu =0,
\end{equation}

\begin{eqnarray} \label{C6}
\frac{1}{2}\ddot{H}
&=&
\sum_{\mu\nu} J_{\mu\nu} (\dot{\bi s}_\nu\cdot\dot{\bi s}_\mu
+  {\bi s}_\mu\cdot\ddot{\bi s}_\nu)\\
&=&
j_{\small min}   \sum_\mu \dot{\bi s}_\mu^2
+ \chi \dot{\bi S}^2 +
\sum_\mu\ddot{\kappa}_\mu +j_{\small min} \sum_\mu
\ddot{\bi s}_\mu\cdot {\bi s}_\mu\\  \label{C6a}
&=&
\sum_\mu\ddot{\kappa}_\mu   + \chi \dot{\bi S}^2,
\end{eqnarray}
using (\ref{C0}) and (\ref{C2}). On the other hand,
\begin{equation} \label{C7}
{H(t)}=\sum_{\mu} {\kappa}_\mu(t)+\chi(t) S^2(t),
\end{equation}
\begin{equation} \label{C8}
\dot{H}(t)=\sum_{\mu} \dot{\kappa}_\mu(t)+\dot\chi(t) S^2(t)
+2 \chi(t)S(t)\dot{S}(t),
\end{equation}
and
\begin{equation} \label{C9}
\ddot{H}=\sum_{\mu} \ddot{\kappa}_\mu
+2 \chi \dot{S}^2.
\end{equation}
Comparison with  (\ref{C6a}) yields $\sum_\mu\ddot{\kappa}_\mu=0$ and
\begin{equation}\label{C10}
\ddot{H}=2\chi\dot{S}^2.
\end{equation}
Hence, using the assumption of parabolicity and l'Hospital's rule,
\begin{equation}\label{C11}
\frac{j-j_{\small min}}{N}=
\left.\frac{dH}{d(S^2)}\right|_{t=0}=
\frac{\ddot{H}}{\frac{d^2}{dt^2}S^2}=
\frac{2 \chi\dot{S}^2}{2 \dot{S}^2}=\chi.
\end{equation}
Summing (\ref{C2}) over $\mu$ yields
\begin{equation}\label{C12}
j \dot{\bi S}= \sum_\mu \dot{\kappa}_\mu {\bi s}_\mu + j_{\small min} \dot{\bi S}+
N \chi \dot{\bi S},
\end{equation}
whence, by (\ref{C11}) and the assumption $\dot{S}\neq 0$,
\begin{equation}\label{C13}
\sum_\mu \dot{\kappa}_\mu {\bi s}_\mu=0.
\end{equation}
We rewrite (\ref{C2}) in the form
\begin{equation}\label{C14}
\sum_\nu (J_{\mu\nu}-j_{\small min}\delta_{\mu\nu}-N\chi E_{\mu\nu})
\dot{\bi s}_\nu =
\dot{\kappa}_\mu {\bi s}_\mu,
\end{equation}
where $E$ is the projector onto ${\Bbb J}$'s eigenvector ${\bi 1}$
with constant entries $E_{\mu\nu}=\frac{1}{N}$. We expand
both sides of (\ref{C14}) into real eigenvectors of ${\Bbb J}$ and obtain
\begin{eqnarray}\label{C15}
\dot{\kappa}_\mu{\bi s}_\mu
&=&
\sum_\alpha{}^\prime
{\bi K}_\alpha C^\mu_\alpha \\  \label{C15a}
\dot{\bi s}_\nu
&=&
\sum_\beta{}^\prime
\frac{1}{j_\beta-j}
{\bi K}_\beta C^\mu_\beta +\frac{1}{N}\dot{\bi S}.
\end{eqnarray}
Now $\sum_\mu \dot{\kappa}_\mu {\bi s}_\mu\cdot\dot{\bi s}_\mu=0$,
$\frac{1}{N}\dot{\bi S}\cdot\sum_\mu \dot{\kappa}_\mu {\bi s}_\mu=0$,
and  $\sum_\mu C^\mu_\alpha C^\mu_\beta = \delta_{\alpha\beta}$
imply
\begin{equation}\label{C16}
0=\sum_\mu
\sum_\alpha{}^\prime {\bi K}_\alpha C^\mu_\alpha
\cdot
\sum_\beta{}^\prime \frac{1}{j_\beta-j}{\bi K}_\beta C^\mu_\beta
=
\sum_\alpha{}^\prime \frac{{\bi K}_\alpha^2}{j_\alpha-j}.
\end{equation}
Since $j>j_\alpha$ for $\alpha>0$ and connected AF systems,
this is only possible if all ${\bi K}_\alpha={\bi 0}$.
Hence, by (\ref{C15a}) $\dot{\bi s}_\nu  = \frac{1}{N}\dot{\bi S}$ and
${\bi s}_\nu \cdot \frac{1}{N}\dot{\bi S}=0$. Thus ${\bi s}$ is coplanar.
\hspace*{\fill}\rule{3mm}{3mm}

An analogous theorem concerning $E_{\small max}(S)$ and coplanar anti-ground
states can be proved similarly.

\section{Extension of ground states}

We consider spin systems with $J_{\mu\nu}=0$ or $J$ and which can thus essentially
be characterized by their graph $\Gamma= ({\cal V},{\cal E})$. A sub-graph
(representing a sub-system)
$\widetilde{\Gamma}= (\widetilde{\cal V},\widetilde{\cal E})$ is defined by the conditions
$\widetilde{\cal V}\subset{\cal V}, \widetilde{\cal E}\subset{\cal E}$ and
$\widetilde{\cal E}\subset \left[\widetilde{\cal V}\right]^2$. Let
\begin{equation}\label{3.1}
\widetilde{H}_0 = 2 J \sum_{(\mu,\nu)\in\widetilde{\cal E}} {\bi s}_\mu \cdot{\bi s}_\nu
\end{equation}
be the Hamiltonian of the subsystem. Let ${\bi s}$ be any state of $\Gamma$ and
$\widetilde{\bi s}$ be its restriction to $\widetilde{\Gamma}$. If ${\bi s}$
happens to be a ground state of $\Gamma$,
$\widetilde{\bi s}$ need not be a ground state of $\widetilde{\Gamma}$, but,
of course, $\widetilde{H}_0(\widetilde{\bi s}) \ge \widetilde{E}_{\small{min}}$. \\

The inverse problem is the extension problem:  Given a ground state $\widetilde{\bi s}$
of  $\widetilde{\Gamma}$, can it be extended to a ground state ${\bi s}$ of $\Gamma$
(such that $\widetilde{\bi s}$ is the restriction of ${\bi s}$)? In general, this
will not be possible. For example, if $\widetilde{\Gamma}$ is the graph of a dimer
with ground state $\widetilde{\bi s}= \uparrow \downarrow $, the extension
of $\widetilde{\bi s}$ to a global ground state is only possible for bi-partite systems.
We have no general theory describing the obstacles against
extension (which could be called frustration of higher order),
but we can describe the typical situation where extension is possible.
For an analogous treatment in the quantum case, see \cite{BSS}. \\

The situation
we have in mind is one where the graph of the spin system $\Gamma_1$ is decomposed
into smaller graphs which are copies of some graph $\Gamma_2$ in such a way that
each edge lies in exactly one of the copies. For example, an icosidodecahedron
can be decomposed into $20$ triangles with disjoint edges. We have local
states by considering the ground states of the copies of $\Gamma_2$. But these
local states must fit together in order make it possible
to construct a global ground state of
$\Gamma_1$. For the example of decomposition into triangles the condition
of fitting-together of local ground states
is equivalent to the $3$-colorability of the large system.
The icosidodecahedron happens to satisfy this condition and hence possesses a
ground state of the energy which is $20$ times the ground state energy of the
triangle. For general $\Gamma_2$ the colorability condition has to be
replaced by the existence of a map from  $\Gamma_1$ to  $\Gamma_2$ which
maps edges onto edges. Such a map may be called a ``graph homomorphism".
\\

A \underline{graph homomorphism} $h: \Gamma_1 \longrightarrow \Gamma_2$ is
a pair of maps $h=(h_{\cal V},h_{\cal E})$ such that
$h_{\cal V}: {\cal V}_1 \longrightarrow {\cal V}_2$,
$h_{\cal E}: {\cal E}_1 \longrightarrow {\cal E}_2$ and
$\{\mu,\nu\}\in {\cal E}_1$ implies
$h_{\cal E}\{\mu,\nu\}=\{h_{\cal V}(\mu),h_{\cal V}(\nu)\}$.
The definition of a \underline{graph isomomorphism} is analogous.

\begin{prop}\label{P3}
Consider two spin systems with graphs $\Gamma_1$ and $\Gamma_2$
and the following properties
\begin{enumerate}
\item
there exists a graph homomorphism $h: \Gamma_1 \longrightarrow \Gamma_2$,
\item there exists a decomposition into sub-graphs $\Gamma_1 =\bigcup_{\mu=1}^{k}{\Gamma_\mu}$,
which is a disjoint decomposition with respect to edges,
\item the restriction of $h$ to $\Gamma_\mu$, $h_\mu:\Gamma_\mu \longrightarrow \Gamma_2$
is a graph isomorphism for each $\mu=1\ldots k$.
\end{enumerate}
Moreover, let ${\bi s}^{(2)}: {\cal V}_2 \longrightarrow {\cal S}^2$
be a ground state of $\Gamma_2$.\\
Then ${\bi s}^{(1)}\equiv h_{\cal V}\circ{\bi s}^{(2)}$ will be a ground state
of $\Gamma_1$.
\end{prop}
{\bf Proof}: According to the disjoint decomposition of ${\cal E}_1$, $H_1$ will
be a sum of terms $H_\mu$ which are each minimized by the state ${\bi s}^{(1)}$. Hence
${\bi s}^{(1)}$ is a ground state of  $\Gamma_1$.
\hspace*{\fill}\rule{3mm}{3mm}  \\

The above-mentioned construction of  ground states for the icosidodecahedron, the octahedron and the
cuboctahedron follows the
description given in this proposition. Other examples show that the two given conditions
of the proposition
are essential: The dodecahedron is $3$-colorable but it cannot be decomposed
into edge-disjoint triangles. Actually its ground state energy is lower than that of the state obtained
by the $3$-coloring. A simpler example is the bi-partite $3$-chain which can be
mapped by a graph homomorphism onto the triangle but is not isomorphic to the triangle.
On the other hand it is possible to connect $4$ triangles such that their edges are
still disjoint but form a tetrahedron which cannot be $3$-colored. This is then a
``higher-order frustrated" system with a ground state energy
which is larger than  $4$ times
the ground state energy of a triangle.


\section{Bounding parabolas}

In this section we will show that the parabolas which occur in theorem \ref{T1}
are general bounds for the extremal values $E_{\small min}(S)$ and
$E_{\small max}(S)$, even if the system is not parabolic. The proof is completely
analogous to the quantum case, see \cite{SSL}.
We recall that $C^{(\nu)}_{\alpha}$ denotes
the $\alpha$-th normalized eigenvector of ${\mathbb J}$
and consider a transformation of the spin
vectors analogous to the transformation onto the eigenbasis
of ${\mathbb J}$. Define
\begin{defi}
${\bi{T}}_\alpha\equiv \sum_\mu \overline{C^\mu_{\alpha}}
{\bi{s}}_\mu,
\mbox{ and  }
{Q}_\alpha\equiv \overline{\bi{T}}_\alpha\cdot {\bi{T}}_\alpha
\quad \alpha=0,\ldots,N-1$.
\end{defi}
The inverse transformation then yields
\begin{equation}\label{B3}
{\bi{s}}_\mu= \sum_\alpha C^\mu_{\alpha}
{\bi{T}}_\alpha,\quad \mu=1,\ldots,N.
\end{equation}
Especially, ${\bi{T}}_{0}={\bi{S}}/\sqrt{N}$ since $\alpha=0$
corresponds to the eigenvector $\bi{1}$.
The following lemma follows directly from the definitions:
\begin{lemma}
\label{BL1}
$
N
=
\sum_{\mu} ({\bi{s}}_\mu)^2
=
\sum_\alpha {Q}_\alpha
=
\frac{1}{N}{\bi{S}}^2
+ \sum'_\alpha {Q}_\alpha
\ .
$
\end{lemma}

Our main result of this section is formulated in the following theorem:
\begin{theorem}\label{BT1}
The following  inequality holds:
\begin{equation}
\frac{j-j_{{min}}}{N} {\bi{S}}^2+j_{{min}} N
\le {H_0} \le
\frac{j-j_{{maxi}}}{N} {\bi{S}}^2+j_{{maxi}} N
\ .
\end {equation}

\end{theorem}

{\bf Proof:} We rewrite the Hamilton function in the following form and conclude
\begin{eqnarray}
\label{B6}
{H_0}
& = &
\sum_{\mu\nu\alpha\beta} J_{\mu\nu} \overline{C^\mu_{\alpha}}C^\nu_{\beta}
 \overline{\bi{T}}_\alpha\cdot {\bi{T}}_\beta
=   \sum_\beta j_\beta {Q}_\beta
=  \frac{j}{N} {\bi{S}}^2 +  \sum_\beta{}^\prime\; j_\beta
{Q}_\beta
\\ \nonumber
& \ge & \frac{j}{N} {\bi{S}}^2 +  j_{{min}} \sum_\beta{}^\prime\;
{Q}_\beta
\\ \nonumber
& = &
\frac{j}{N} {\bi{S}}^2 +j_{{min}}\left( N -\frac{1}{N}{\bi{S}}^2 \right)
=
\frac{j-j_{{min}}}{N}{\bi{S}}^2 +j_{{min}} N
\ ,
\nonumber
\end {eqnarray}
using (\ref{B3}), the positivity of ${Q}_\beta$, and lemma \ref{BL1}.
The other inequality follows
analogously.\hspace*{\fill}\rule{3mm}{3mm}  \\

We note that the proof does not depend on the dimension of spin space, but the optimal
energy bounds may depend on this dimension. The regular pentagon
is an example of a system where the above bounding parabolas are
assumed by $3$-dimensional states, but not by $2$-dimensional, i.~e.~, coplanar states, see
section 6.3.1.

\section{Symmetric states}
In this section we will assume suitable symmetry properties of the spin system and
correspondingly consider the notion of ``symmetric states". We make use of some
simple concepts and results of group theory, which may be found in many textbooks, e.~g.~
\cite{LF}.\\

Let $J_{\mu\nu}=0$ or $J$ and consider the graph $\Gamma=({\cal {V,E}})$
characterizing the spin system.
Further let ${\cal G}$ denote the symmetry group of the graph $\Gamma$.
Hence ${\cal G}$ consists of all permutations of ${\cal V}$ which map
edges onto edges. ${\cal G}$ has a ``natural" representation by real $N\times N$-matrices,
by permuting the standard basis of ${\Bbb R}^N$ in the same way as the spin sites
$\mu\in{\cal V}$. We will denote the  $N\times N$-matrix representing any symmetry
$g\in{\cal G}$ by ${\bf g}$ and by ${\Bbb G}$ the set of all  $N\times N$-matrices obtained
in this way. Obviously,
\begin{lemma}\label{L3}
The adjacency matrix ${\Bbb J}$ commutes with all ${\bf g}\in{\Bbb G}$.
\end{lemma}
Hence
\begin{lemma}\label{L4}
The eigenspaces of ${\Bbb J}$ split into orthogonal direct sums of irreducible
subrepresentations of the natural representation of ${\cal G}$.
\end{lemma}
The irreducible representations of the relevant groups are well
known, see e.~g.~ \cite{LF}, and can easily be associated to the different
(subspaces of) eigenspaces of ${\Bbb J}$.\\

For specific spin systems, e.~g.~magnetic molecules, the spin sites
$\mu\in{\cal V}$ are embedded into the physical 3-space and the group ${\cal G}$
could also be identified with one of the point groups,
i.~e.~finite subgroups of ${\cal O}(3,{\Bbb R})$.
There are only a finite number of possibilities, see e.~g.~\cite{LF}, 3.1.2. and 3.1.3.
The most interesting cases are
\begin{itemize}
\item the dihedral group ${\cal D}_n$,
\item the tetrahedral group ${\cal T}$,
\item the octahedral group ${\cal O}$,
\item the icosahedral group ${\cal Y}$,
\end{itemize}
as well as the improper point groups attached to them.\\

In what follows, we make the following general assumption:
\begin{ass}
All exchange parameters $J_{\mu\nu}$ are $0$ or $J$, and the group ${\cal G}$ operates
transitively on ${\cal V}$.
\end{ass}
The latter means that for each pair $\mu,\nu\in{\cal V}$ there is a permutation
$g\in{\cal G}$ such that  $g(\mu)=\nu$. The two groups ${\cal G}$ and
${\cal O}(3,{\Bbb R})$ operate independently on the set of states: If two matrices
${\bf g}\in{\Bbb G}$ and $R\in{\cal O}(3,{\Bbb R})$ are given, we define their
action on a state $\bi{s}$ by matrix multiplication with  $\bi{s}$ which
is again considered as an $N\times 3$-matrix:
\begin{defi}
$(R,{\bf g})\bullet \bi{s} \equiv R \bi{s} \bf{g}^{-1}.$
\end{defi}

The transformations $R\in{\cal O}(3,{\Bbb R})$ will also be referred to as
``rotations/reflections in spin space".\\

We have the following obvious results:
\begin{lemma}\label{L5}
Let ${\bf g}\in{\Bbb G}$ and $R\in{\cal O}(3,{\Bbb R})$ and $\bi{s}$ be a state. Then
\begin{enumerate}
\item  $\bi{s}$ and $ R \bi{s} \bf{g}^{-1} $  have the same total spin length $S$,
\item If $\bi{s}$ is (weakly) stationary, then $R \bi{s} \bf{g}^{-1}$ is
(weakly) stationary,
\item   If $\bi{s}$ is weakly symmetric, then
$R \bi{s} \bf{g}^{-1}$ is  weakly symmetric,
\item if $\bi{s}$ is a (anti-) ground state, then
$R \bi{s} \bf{g}^{-1}$ is a (anti-) ground state.
\end{enumerate}
\end{lemma}

Having defined the action of the product group $ {\cal O}(3,{\Bbb R})\times {\Bbb G}$
on states,
it is natural to consider the corresponding subgroup leaving a given state invariant:

\begin{defi}
For any state $\bi{s}$ let ${\cal G}_\bi{s}\equiv
\{ (R,{\bf g})\in {\cal O}(3,{\Bbb R})\times {\Bbb G} | R \bi{s} {\bf g}^{-1} = \bi{s}\}.
{\quad\cal G}_\bi{s}$ is called the
\underline{symmetry group} of $\bi{s}$.
\end{defi}

\begin{defi}
A state  $\bi{s}$ is called \underline{symmetric} if
the projection onto the second factor
$\pi_2: {\cal G}_\bi{s} \longrightarrow {\Bbb G}$ is surjective,
i.~e.~if for each ${\bf g}\in{\Bbb G}$ there exists a rotation/reflection
$R\in{\cal O}(3,{\Bbb R})$ such that $R^{-1}\bi{s}{\bf g}=\bi{s}$.
\end{defi}

One may thus say that for symmetric states $\bi{s}$ every permutation
${\bf g}\in{\Bbb G}$ of the vectors $\bi{s}_\mu$ can be compensated by a
suitable rotation/reflection in spin space.\\

For example, if $({\cal V},{\cal E})$ is the graph of a regular $N$-polygon,
the coplanar symmetric states are in $1:1$-correspondence to the roots of
unity, $z_n^N=1$, namely
\begin{equation}\label{13}
\bi{s}_\mu=z_n^\mu=\exp(i\mu n 2 \pi/N),\quad n, \mu=1,\ldots,N.
\end{equation}
Here ${\cal G}={\cal D}_N$, the dihedral group of order $2N$.\\

We have the following
\begin{lemma}\label{L6}
Let $\bi{s}_{(\mu)}^{(i)}$ be a non-vanishing $N\times 3$-matrix such that
for all $\bi{g}\in{\Bbb G}$ there exists some
$R\in{\cal O}(3,{\Bbb R})$ such that $\bi{s}{\bf g}=R\bi{s}$.
Then, for some suitable $\lambda\in{\Bbb R}$,
$\lambda\bi{s}$ will be a symmetric state.

\end{lemma}
{\bf Proof:}
It remains to show that $\lambda\bi{s}_\mu^{(i)}$ will be unit vectors
for all $\mu=1,\ldots,N$. Fix some indices $\mu,\nu$.
Using assumption 1, we choose a $\bi{g}\in{\Bbb G}$ such that
$(\bi{s g})_\mu = \bi{s}_\nu = R \bi{s}_\mu$. Since
$R\in{\cal O}(3,{\Bbb R})$  the two vectors
$\bi{s}_\nu$ and $\bi{s}_\mu$  have the same length.
Because  $\mu,\nu$ were arbitrarily chosen, all vectors
$\bi{s}_\mu$ have the same length, say $\lambda^{-1}$,
which completes the proof.
\hspace*{\fill}\rule{3mm}{3mm}  \\

If a state $\bi{s}$ is not collinear, then $R \bi{s} = \bi{s}$ implies $R=\Eins$.
In this case for each ${\bf g}\in \pi_2({\cal G}_\bi{s})$ there exists a unique
$R\in{\cal O}(3,{\Bbb R})$ such that $R^{-1}\bi{s}{\bf g}=\bi{s}$. We will write
$R=\rho_\bi{s}({\bf g})$. It is easily shown that the map
${\bf g}\mapsto \rho_\bi{s}({\bf g})$  is a linear representation of the subgroup
$\pi_2({\cal G}_\bi{s})\subset {\Bbb G}$. If $\bi{s}$ is  collinear we may uniquely
fix $R$ by choosing $R\in\{\Eins,-\Eins\}$
and thus obtain a representation $\rho_\bi{s}$ also in this case.\\

For the case of symmetric $\bi{s}$ we have   $\pi_2({\cal G}_\bi{s})= {\Bbb G}$
and hence  $\rho_\bi{s}$ will be an  $n$-dimensional representation of ${\Bbb G}$
with $n\le 3$.

In the case of a symmetric ground state it is thus necessary that the eigenspace
of ${\Bbb J}$ corresponding to the smallest eigenvalue $j_{{\small min}}$ contains an
irreducible representation of  ${\Bbb G}$ of dimension $n\le 3$. If it does not,
we can exclude symmetric ground states.\\

For practical purposes of constructing (anti-)ground states
it would be desirable to invert the process and to reconstruct
$\bi{s}$ from a given $n$-dimensional subrepresentation $\rho$ of the natural
representation of ${\cal G}$, $n\le 3$ .
We will now describe this procedure and consider, for sake of simplicity, only the
case $n=3$. Let ${\cal S}$ be an $3$-dimensional subspace of ${\Bbb R}^N$ left
invariant by all ${\bi g}\in {\Bbb G}$ and ${\bi s}_{(\mu)}^i,\; i=1,2,3$ a basis
of ${\cal S}$.  It follows that the three vectors
$\sum_\mu {\bi g}_{\mu(\nu)}\bi{s}_\mu^i,\; i=1,2,3$ are linear combinations
of the basis vectors, hence
\begin{equation}
\sum_\mu {\bi g}_{\mu\nu}{\bi s}_\mu^i
=
\sum_j \rho_j^i({\bi g}){\bi s}_\nu^j,
\end{equation}
or, in matrix notation, $\rho({\bi g}){\bi s} = \bi{s g}$.
Being a real representation of a finite group,
$\rho$ is equivalent to an orthogonal representation,
i.~e.~$\widetilde{\rho}({\bi g})\equiv T \rho({\bi g}) T^{-1}$
is orthogonal with some invertible $3\times 3$-matrix $T$. It follows that
\begin{equation}
\widetilde{\rho}({\bi g}) (T{\bi s}) = (T{\bi s}) {\bi g},
\end{equation}
which means that $T{\bi s}$ is a symmetric state with
$\rho_{T{\bi s}}=\rho$. Note that the orthogonality of the representation
is crucial in order to obtain a row
$(T{\bi s})_\mu^{(i)}\; \mu=1,\ldots,N$
of unit vectors, compare lemma \ref{L6}. Hence we have proven the following

\begin{prop}\label{P4}
Let $\rho$ be a $3$-dimensional subrepresentation of the natural
representation of ${\cal G}$. Then there exists a $3$-dimensional
symmetrical state ${\bi s}$ such that  $\rho_{{\bi s}}=\rho$.
\end{prop}

Further there holds
\begin{prop}\label{P1}
Each stationary symmetric state  $\bi{s}$ is weakly symmetric.
\end{prop}
{\bf Proof:} We will use a matrix notation and write
$\bi{\kappa}\equiv{diag}(\kappa_1,\ldots,\kappa_N)$. Then the condition (\ref{8})
for stationarity may be written as $\bi{s}{\Bbb J}=\bi{s}\bi{\kappa}$.
For ${\bf g}\in{\Bbb G}$ we have
$\widetilde{\kappa}\equiv {\bf g}^{-1}\kappa{\bf g} =
{diag}(\kappa_{{\bf g}(1)},\ldots,\kappa_{{\bf g}(N)})$.
We conclude $\bi{s}\kappa {\bf g} = \bi{s}{\Bbb J} {\bf g} = \bi{s}{\bf g}{\Bbb J}
=\rho_{\bi s}({\bf g})\bi{s}{\Bbb J} = \rho_{\bi s}({\bf g})\bi{s}\kappa
= \bi{s}{\bf g}({\bf g}^{-1}\kappa {\bf g}) = \rho_{\bi s}({\bf g})\bi{s}\widetilde{\kappa}.$
Hence $\bi{s}\kappa=\bi{s}\widetilde{\kappa}$, which is in components
$\kappa_\mu \bi{s}_\mu=\kappa_{{\bf g}(\mu)} \bi{s}_\mu$ whence $\kappa_\mu$ is
independent of $\mu$ (using assumption 1).
\hspace*{\fill}\rule{3mm}{3mm}  \\

The converse of proposition \ref{P1} is not true: There are ground states of the
icosidodecahedron which are weakly symmetric, but not symmetric, see section 6.5.
The condition of stationarity in proposition \ref{P1} can be replaced by
another condition:

\begin{defi}\label{15}
Let ${\cal G}_{\bi s}$ denote the subgroup of ${\cal G}$ leaving
$\mu\in{\cal V}$ fixed and  ${\Bbb G}(\mu)$ the matrix group
generated by its natural representation.
A symmetric state $\bi{s}$ will be called \underline{isotropic} if
$\rho_{\bi{s}}({\Bbb G}(\mu))$
contains at least one rotation with an angle $\alpha\neq 0,\pi$.
\end{defi}

\begin{lemma}\label{L7}
If $ \bi{s}$ is isotropic and $\rho_{\bi{s}}=\rho_{\bi{s'}}$
then $\bi{s}_\mu =\pm \bi{s'}_\mu$.
\end{lemma}
{\bf Proof:} $\bi{s}_\mu$ is invariant under all rotations/reflections
$\rho_{\bi{s}}({\bf g}), {\bf g}\in{\Bbb G}(\mu)$. Since some of these is
a rotation with an angle $\alpha\neq 0,\pi$, the axis of rotation will be
unique and $\rho_{\bi{s}}=\rho_{\bi{s'}}$ implies   $\bi{s}_\mu =\pm \bi{s'}_\mu$.
\hspace*{\fill}\rule{3mm}{3mm}  \\

\begin{prop}\label{P2}
Each isotropic state is stationary, hence weakly symmetric.
\end{prop}
{\bf Proof:} Using the above matrix notation we obtain
$\bi{s}{\Bbb J}{\bf g}=\bi{s}{\bf g}{\Bbb J}= \rho_{\bi{s}}({\bf g})\bi{s}{\Bbb J}$
hence $\bi{s}{\Bbb J}$ is a symmetric (not normalized) state with
$\rho_{\bi{s}{\Bbb J}}=\rho_{\bi{s}}$. Using that $\bi{s}$ is isotropic and lemma (\ref{L7})
we conclude $(\bi{s}{\Bbb J})_\mu =\pm\lambda\bi{s}_\mu$, hence $\bi{s}$ is stationary,
and, by proposition (\ref{P1}), weakly symmetric. \hspace*{\fill}\rule{3mm}{3mm}  \\

\section{Examples}
We will mainly consider AF systems and set $J>0$ throughout this section
if not mentioned otherwise.
\subsection{The general spin triangle}

We consider spin systems with $N=3$ and general
coupling coefficients $J_1, J_2$ and $J_3$. The special symmetric case
$J_1=J_2=J_3$ is atypical and its properties are discussed later.
Another special case is the $3$-chain with $J_1=0, J_2=J_3$, which is
probably the simplest example of a non-parabolic system. Its investigation will be left
for the reader.
\\

\begin{figure}[f01]
\begin{center}
\epsfig{file=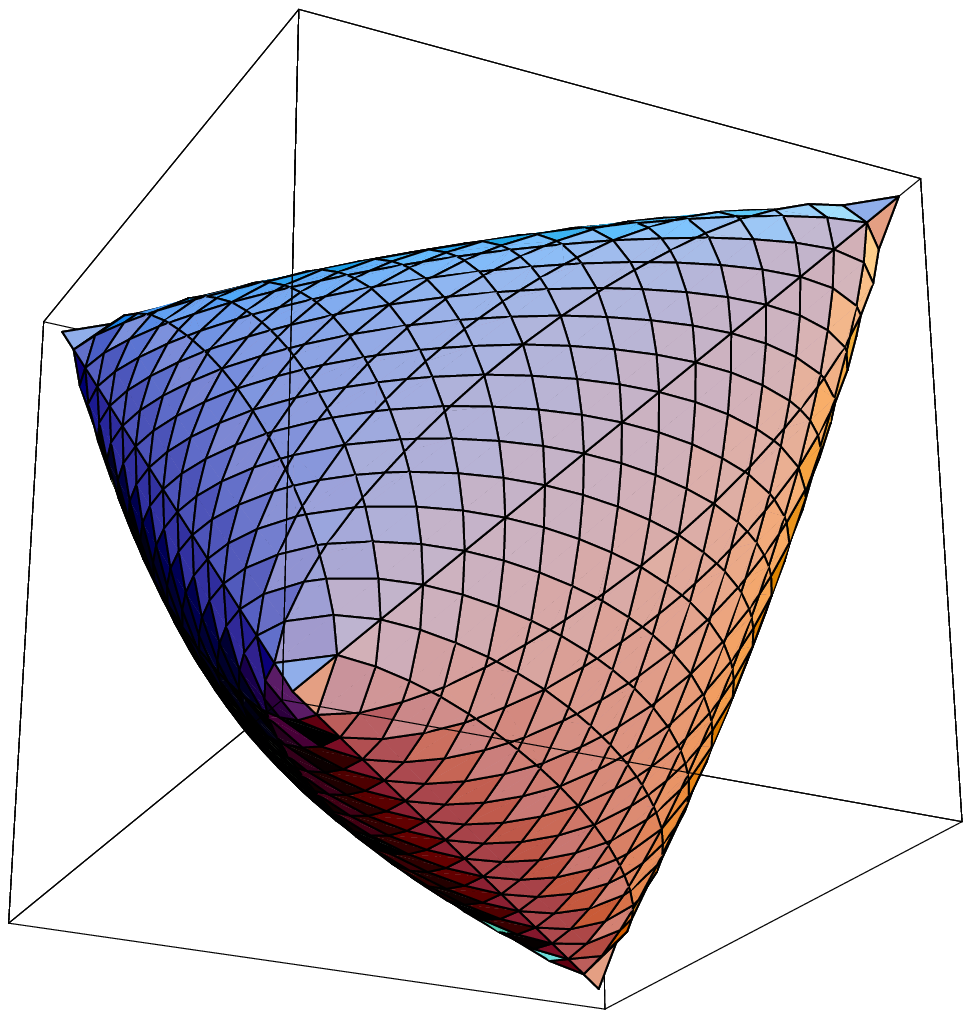,width=100mm}
\vspace*{1mm}
\caption[]{The $3$-dimensional convex set ${\cal P}$ defined in (\ref{5.2})
as seen from the view point $(4,-2,-2)$. Its shape is essentially identical with
that shown in
figure \ref{F-2}.}
\label{F-1}
\end{center}
\end{figure}

\begin{figure}[f02]
\begin{center}
\epsfig{file=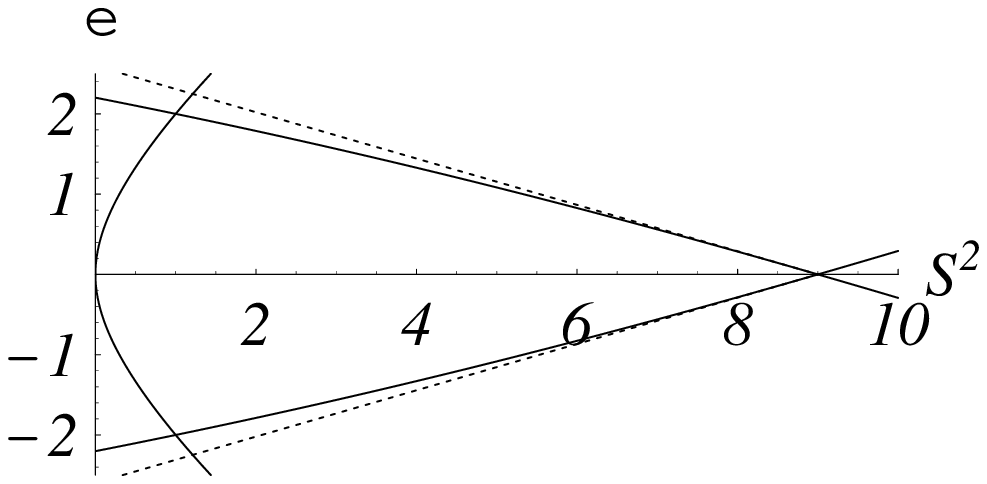,width=150mm}
\vspace*{1mm}
\caption[]{Energy $\epsilon$ versus square of total spin, $S^2$,
for the triangle with $J_1=0,\, J_2=1,\, J_3=-1$ according
to (\ref{5.9}). The broken lines
represent the bounding parabolas of theorem \ref{BT1}.}
\label{F-2}
\end{center}
\end{figure}

Let ${\bf s}_\nu, \nu=1,2,3,$ be any spin configuration and consider the $3\times 3$-matrix
${\cal S}$ with coefficients
\begin{equation}\label{5.1}
{\cal S}_{\mu\nu} = {\bf s}_\mu \cdot{\bf s}_\nu.
\end{equation}
It is positive and its diagonal consists of $1$'s. Conversely, any $3\times 3$-matrix
with these properties is of the form (\ref{5.1}). This follows from the spectral theorem.
These matrices can be written in the form
\begin{equation}\label{5.2}
{\cal S} = \left(\begin{array}{ccc}1&w&v\\w&1&u\\v&u&1\end{array}\right) \equiv [u,v,w],
\end{equation}
where $u,v,w$ are real numbers subject to the constraint
\begin{equation}\label{5.3}
\det{\cal S} =1-(u^2 +v^2+w^2)+2 u v w \ge 0, \; \mbox{and } u^2, v^2, w^2 \le 1.
\end{equation}
Hence the set ${\cal P}$ of the matrices of the form (\ref{5.1}) can be considered
as a compact convex subset of ${\mathbb R}^3$. It has the form of an ``inflated tetrahedron",
see figure \ref{F-1}.
 Since the map
\begin{eqnarray}\label{5.4}
&\pi:& {\cal P} \longrightarrow   {\mathbb R}^2 \\
\pi[u,v,w] &\equiv& (3+2(u+v+w), J_1 u + J_2 v + J_3 w) \\
&=& (S^2({\bf s}),\frac{1}{2} H_0({\bf s}) )
\end{eqnarray}
is affine, the set ${\cal P}_2\equiv \pi[{\cal P}]$ of all possible values of the
energy vs.~square of the total spin will be a compact convex
set too. To simplify the geometry we will assume $J_1+J_2+J_3=0$.
Then  ${\cal P}_2$ can be considered as the orthogonal projection of
${\cal P}$ onto a suitable plane. If we introduce new coordinates in the
$u,v,w$-space by
\begin{eqnarray}\label{5.5}
\sigma &=& u+v+w\\
\epsilon &=& J_1 u + J_2 v + J_3 w \\
\tau &=&(J_2-J_3)u+ (J_3-J_1)v+(J_1-J_2)w
\end{eqnarray}
the orthogonal projection $\pi$ is essentially the projection onto the
$\sigma,\epsilon$-plane. Hence the boundary of ${\cal P}_2$ can be obtained
by solving the two equations
\begin{eqnarray}\label{5.6}
\det{\cal S} &=& 0,\\  \label{5.7}
\nabla \det{\cal S} \cdot \left(\begin{array}{c}J_2-J_3\\ J_3-J_1\\J_1-J_2\end{array}\right)
&=&0 .
\end{eqnarray}
The first equation (\ref{5.6}) defines the boundary of ${\cal P}$, the second one (\ref{5.7})
is satisfied by those points of the  boundary of ${\cal P}$ which are orthogonally projected
onto the boundary of ${\cal P}_2$. By using computer algebra software it is straightforward
to express (\ref{5.6}) and (\ref{5.7}) as equations for $\sigma,\epsilon,\tau$ and to
eliminate $\tau$ between  (\ref{5.6}) and (\ref{5.7}). Yet the result for the general case
is too complex to be reproduced here. We choose the special case
\begin{equation}\label{5.8}
J_1=0, J_2=1, J_3=-1
\end{equation}
and obtain the following equation for the boundary of ${\cal P}_2$
\begin{eqnarray}\nonumber
0&=&
S^2(9 - 10 S^2 + S^4)^2 +(-27 + 288 S^2 + 18 S^4 - 24 S^6 + S^8)\epsilon^2 \\
\label{5.9}
&&
-8(3 + S^2)^2 \epsilon^4 + 16\epsilon^6 ,
\end{eqnarray}
where we have re-substituted $\frac{S^2-3}{2}$ for $\sigma$.  The actual boundary
of ${\cal P}_2$ is only a part of the family of curves defined by (\ref{5.9}) and is shown in figure
\ref{F-2}. Here we have an example where the curves $E_{\small{min}}{(S)}$ and $E_{\small{max}}{(S)}$
can be explicitely calculated and are not parabolas, although they are close to their
bounding parabolas which are represented as broken lines in figure \ref{F-2}.
This is not
in contradiction with theorem \ref{T1} since the ground state of this system is
coplanar but has $S>0$. \\

We expect that similar figures also appear for the general system of $N$ spins.
Parabolic
systems are probably rare cases, some of which will be considered in the remaining
subsections.


\subsection{${\cal S}_N$-symmetric systems}

Recall that the  $N$-pantahedron was defined as a system of $N$ spins where each
pair $(\mu,\nu)$ interacts with equal strength $J>0$. These systems serve mainly as
exactly solvable model examples for the sake of illustration, but the dimer, the triangle,
and the tetrahedron are also of experimental relevance
\cite{JMMMS},\cite{MMBSB},\cite{FLBJMMBMS}.
The symmetry group of the $N$-pantahedron is obviously the group  ${\cal S}_N$
of all permutations of $N$ spin sites. \\

By completing squares the Hamiltonian can be written as
\begin{equation} \label{16}
H_0=J(\bi{S}^2 - N).
\end{equation}
Since the energy depends only on $S$, each state is a relative (anti-)ground state
with
\begin{equation} \label{17}
E_{{\small min}}(S)=E_{max}(S)=J(S^2-N).
\end{equation}

For $S=0$ there exist symmetrical ground states only if $N\le 4$. These
are the obvious dimer, triangle, and tetrahedron configurations in spin space.
The dimension $dim$ of the irreducible representations of ${\cal S}_N$
is given by the number of standard Young tableaus (see \cite{LF}) and hence $dim=1$ or
$dim\ge N-1$. The one-dimensional representations are either the trivial one corresponding
to the total anti-ground state or the one attaching to each permutation its sign, which
is not generated by a symmetric state. Thus there are no symmetrical ground states
for $N>4$. For example, a cyclic permutation of $3$ spins leaving $2$ other spins
invariant cannot be compensated by a reflection/rotation in spin space.\\

The coplanar $N$-polygon configuration in spin space does however define a weakly symmetric ground state
with $\kappa_\mu=-1$.
Hence the  construction of theorem 1  yields
symmetrical relative ground states $\hat{\bi{s}}(S)$. We will check our equation
(\ref{11}) for this case:
Let ${\Bbb E}$ denote the $N\times N$-matrix with $E_{\mu\nu}=1/N$ for all $\mu,\nu$.
It is the matrix of the projector onto the one-dimensional subspace
spanned by the vector $\bi{1}=\frac{1}{\sqrt{N}}(1,1,\ldots,1)$. The adjacency matrix for the
$N$-pantahedron is ${\Bbb J}=J(N{\Bbb E}-\Eins)$. Hence its eigenvalues
are $j=J(N-1)$ with eigenvector $\bi{1}$ and $j_{{\small min}}=-J$ with the $(N-1)$-dimensional
eigenspace of vectors orthogonal to $\bi{1}$. Thus $j_{{\small min}}=j_{\small maxi}$ and the two
functions $E_{{\small min}}(S)$ and $E_{max}(S)$ coincide in accordance with what has been said above.

\subsection{${\cal D}_N$-symmetric systems}
\subsubsection{The $N$-polygon}

The $N$-polygon has the dihedral group ${\cal D}_N$ as symmetry group.
We have already mentioned the set of $S=0$, stationary, coplanar, symmetric states $\bi{s}$
given by
\begin{equation}\label{17}
\bi{s}_\mu =z_n^\mu,\quad \mu=0,\ldots,N-1,
\end{equation}
where $z_n=\exp(i n 2 \pi/N), n=1,\ldots,N,$ is a root of unity, $z_n^N=1$.
Indeed, these states are the complex eigenvectors of the corresponding adjacency matrix
\begin{equation}\label{18}
{\Bbb J}= J \left( \begin{array}{ccccccc}
0 & 1 & 0 & \ldots &\ldots &  0 &1 \\
1 & 0 & 1 & 0 & \ldots &\ldots & 0 \\
0 & 1 & 0 & 1 & 0 & \ldots & 0 \\
\vdots & & & & & & \\
0 & \ldots &\ldots &  0 & 1 & 0 & 1\\
1 & 0 & \ldots & \ldots &  0    & 1 & 0
\end{array}\right)
\end{equation}

The eigenvalues are obtained by
\begin{equation}\label{19}
({\Bbb J} \bi{s})_\mu= J (z_n^{\mu+1}+z_n^{\mu-1})=2 J \cos(2\pi n/N) \bi{s}_\mu
\end{equation}
and the corresponding energies are
\begin{equation}\label{20}
E_n= 2JN \cos(2\pi n/N),\quad n=0,\ldots,N-1.
\end{equation}
It follows that
\begin{equation}\label{21}
E_{{\small min}}=N j_{{\small min}} =\left\{ \begin{array}{lcll}
E_{N/2} & = & -2JN & \mbox{for } N \mbox{ even,}\\
E_{(N-1)/2} & = & 2JN\cos(\pi (N-1)/N) & \mbox{for } N \mbox{odd}\\
\end{array}\right.
\end{equation}
and
\begin{equation}\label{22}
E_{\small max}=E_0=Nj=2NJ.
\end{equation}
This  follows since the bounds of lemma 1 are attained by these states.
For $N=5$ the ground state of the pentagon may be vizualized as a pentagram
in spin space and analogously for other odd $N$.\\

Moreover,
\begin{equation}\label{23}
E_{\small max}(0)=E_1=Nj_{\small maxi}=2NJ\cos(2\pi/N).
\end{equation}
The relative ground states of theorem 1 have energies
\begin{equation}\label{24}
E_{{\small min}}(S)=\left\{ \begin{array}{ll}
2J(2 \frac{S^2}{N}-N) & \mbox{for} N \mbox{even,}\\
2J((1-\cos(\pi \frac{N-1}{N})) \frac{S^2}{N}+N \cos(\pi\frac{N-1}{N}))
& \mbox{for} N \mbox{odd},\\
\end{array}\right.
\end{equation}
and
\begin{equation}\label{25}
E_{max}(S)=2J((1-\cos(2\pi/N)\frac{S^2}{N}+N\cos(2\pi/N)).
\end{equation}
Both parabolas meet in the anti-ground state with $E_{\small max}= 2JN$ for $S=N$.

\subsubsection{Coplanar states of the pentagon}

\begin{figure}[f07]
\begin{center}
\epsfig{file=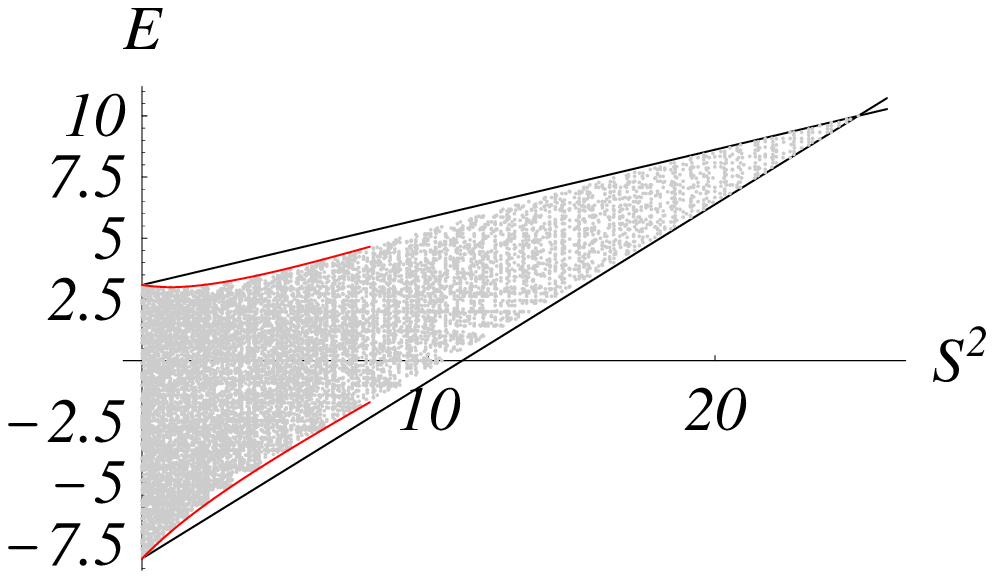,width=150mm}
\vspace*{1mm}
\caption[]{The shaded region represents the spectrum $E$ versus $S^2$ of coplanar states
of the pentagon as determined by numerical methods. The straight lines are the boundaries of the
full spectrum. Also a part of the boundary of the coplanar spectrum which can be analytically
calculated is displayed.}
\label{F-7}
\end{center}
\end{figure}

The regular pentagon as a special case dealt with in the preceding subsection assumes its bounding
parabolas. However, the extremal energies $E_{\small min}(S)$ and  $E_{\small max}(S)$
can only be realized by non-coplanar states, except for $S=0$ and $S=5$. This has not been rigorously
proven but shown by numerical evidence, see figure \ref{F-7}.  The spectrum $E$ versus $S^2$ realized
by coplanar states is a subset of the full spectrum with concave boundaries
$E_{\small min}^{\small coplanar}(S)$ and  $E_{\small max}^{\small coplanar}(S)$.

The permutation of spin sites
$(0\leftrightarrow 0,\; 1\rightarrow 2\rightarrow 4\rightarrow 3 \rightarrow 1)$
leaves the Hamiltonian invariant and maps relative ground states onto anti-ground states, hence
$E_{\small min}^{\small coplanar}(S)= S^2 - 5 - E_{\small max}^{\small coplanar}(S)$.
The (anti-) ground state with $S=1$ and $E=-7$ (resp.~$E=3$) is especially remarkable:
Its state vectors ${\bi s}_\mu$ occupy the vertices of a regular hexagon leaving one vertex free.
The anti-ground state ($E_{\small max}^{\small coplanar}(1)=3$) satisfies equation (\ref{8}) with $\chi=0$,
hence it is a stationary state, not only weakly
stationary as other relative anti-ground states. Actually, $E_{\small max}^{\small coplanar}$ has
a local minimum at $S=1$, see figure \ref{F-7}. For axisymmetric weakly stationary
states in the neighborhood of the hexagonal state,
equation (\ref{8}) can be solved analytically
by using computer algebra software and hence
$E$ and $S^2$ can be expressed as functions of a common
parameter, although of forbidding complexity. The resulting curve turns out to be
a part of the upper boundary of the coplanar spectrum and is displayed in figure \ref{F-7},
as well as the corresponding lower boundary part.

\subsubsection{The pentagonal star}

\begin{figure}[f03]
\begin{center}
\epsfig{file=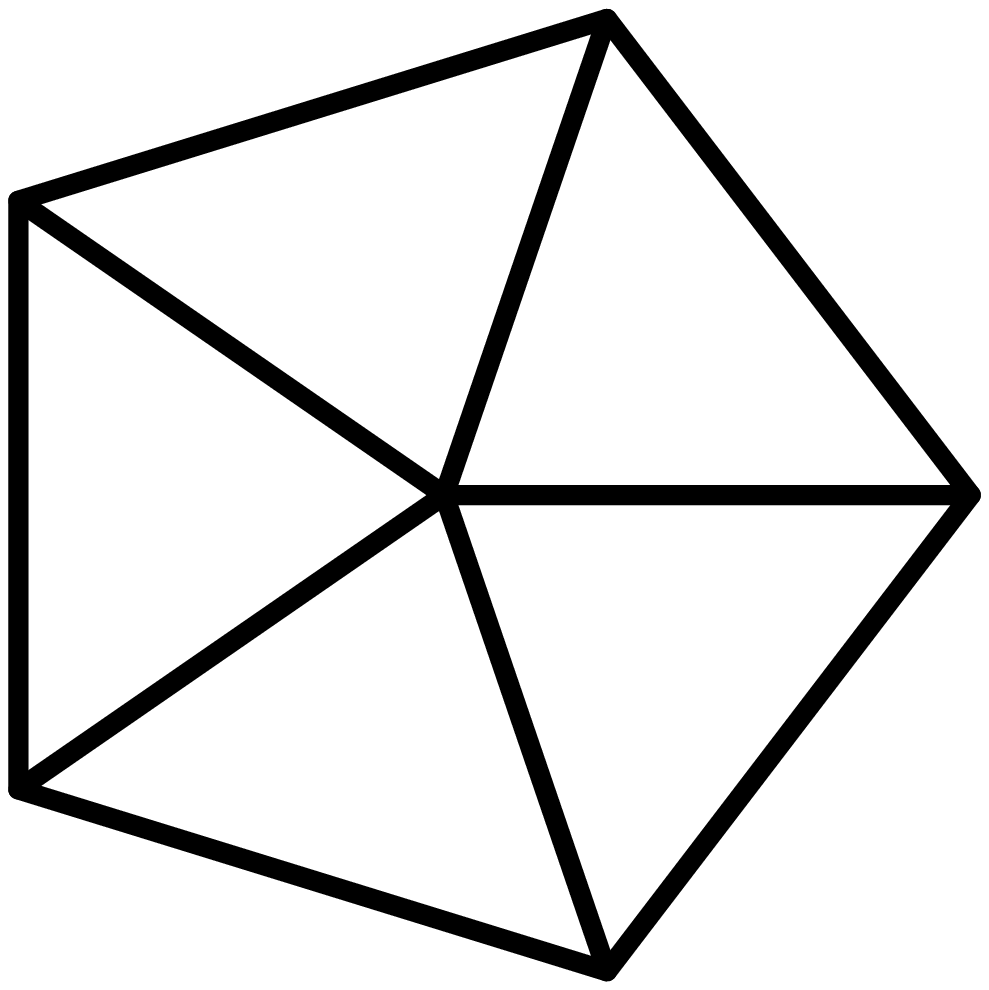,width=70mm}
\vspace*{1mm}
\caption[]{Pentagonal star as an example of a spin system without coplanar ground states.}
\label{F-3}
\end{center}
\end{figure}

We obtain the pentagonal star ($N=6$) by joining the $5$ vertices of the pentagon
with its mid-point, see figure \ref{F-3}. This system has ${\cal D}_5$ symmetry, but
it is not ${\cal D}_5$-symmetric in our sense, since ${\cal D}_5$ does not operate
transitively on the $6$ spin sites. Nevertheless, we will discuss this example since
it has the interesting property that its ground states are not coplanar and
$E_{\small min}(S)$, $E_{\small max}(S)$ are only piecewise parabolic.\\

\begin{figure}[f04]
\begin{center}
\epsfig{file=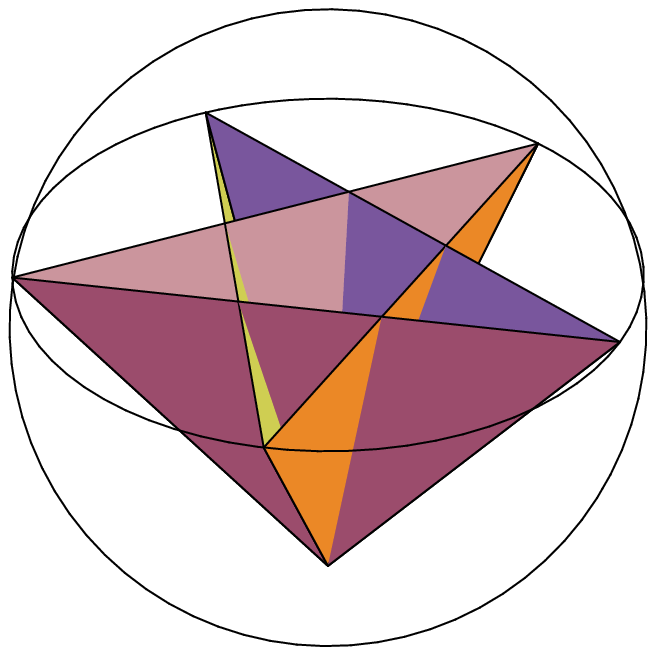,width=100mm}
\vspace*{1mm}
\caption[]{$3$-dimensional ground state of the pentagonal star: The vertices
correspond to $6$ unit vectors in spin space and the
edges correspond to the $10$ bonds of the pentagonal star of figure \ref{F-3},
see
the remarks after (\ref{4f}).}
\label{F-4}
\end{center}
\end{figure}

If we add to the $10$ edges of the pentagonal star the 5 edges of the corresponding
pentagram we obtain the $15$ edges of the $6$-pantahedron. Hence its Hamiltonian reads
\begin{equation}\label{25a}
H= S^2-6-H_5,
\end{equation}
where $H_5$ is the Hamiltonian of the pentagram, which is the same as that for the
pentagon, up to a suitable permutation of the spin sites. This shows that the configurations
minimizing $H$ for given $S$ are exactly those which maximize $H_5$, analogously for
$E_{\small max}(S)$. The maximal values for $H_5$ are given by the parabola, compare (\ref{25}),
\begin{equation}\label{25b}
E_{\small max}^{(5)}(S_5)=\frac{2(1-\cos(2\pi/5))}{5}S_5^2 + 10 \cos(2\pi/5),
\end{equation}
where $S_5$ is the length of the total spin of the $5$ vertices of the pentagon.
Here $S_5$ has to be chosen maximal for given $S$, that means that
$S_5=S+1$ for $0\le S \le 4$ and $S_5=5$ for $4\le S \le 6$. This yields
\begin{equation}\label{25c}
E_{\small min}(S)=
\left\{
\begin{array}{lcl}
\frac{5+\sqrt{5}}{10}(S-4)(S+1+\sqrt{5})
&
\mbox{if}
&
0\le S \le 4\\
S^2-16
&
\mbox{if}
&
4\le S \le 6
\end{array}
\right.
\end{equation}
Hence the total ground state is attained not for $S=0$,
but for $S=\frac{3-\sqrt{5}}{2}=0.381966\ldots$
with $E_{\small min}=-5-2\sqrt{5}=-9.47214\ldots$.
The corresponding spin configuration is shown in figure \ref{F-4}.

It remains to show that no other, coplanar ground state exists.
 Employing the numerical result of the preceding subsection for the pentagon,
$E_{\small max}^{\small coplanar}(S)  < E_{\small max}(S)$
for $0<S<5$,
it is easily shown that all
coplanar states with $S=\frac{3-\sqrt{5}}{2}$ have an energy larger
than $E_{\small min}$. The minimal
energy for all coplanar states appears to be $E_{\small min}^{\small coplanar}=-9$,
realized by a hexagonal ground state with $S=0$, but we do not yet have a
rigorous proof for this claim.
\\

Analogously one can show that
\begin{equation}\label{25d}
E_{\small max}(S)=
\left\{
\begin{array}{lcl}
\frac{5-\sqrt{5}}{10}(S+4)(S-1+\sqrt{5})
&
\mbox{if}
&
1\le S \le 6\\
S^2-16+\frac{5(1+\sqrt{5})}{2}
&
\mbox{if}
&
0\le S \le 1
\end{array}
\right.
\end{equation}

\subsection{${\cal O}$-symmetric systems}

\subsubsection{The cube}

As already mentioned the cube allows a bi-partition and hence possesses a ground state
of the form
\begin{equation}\label{26}
\bi{s}_\mu = (-1)^\mu \bi{e},\quad \mu=1,\ldots,8,
\end{equation}
if the vertices are suitably labelled. From this we obtain relative ground states
with
\begin{equation}\label{27}
E_{{\small min}}(S)=J(\frac{3}{4}S^2-24),
\end{equation}
since $j=3J, j_{{\small min}}=-3J$. Apart from $j$ and $j_{{\small min}}$ the other
eigenvalues of ${\Bbb J}$ are $\pm J$ with three-fold degeneracy, respectively.
The corresponding eigenspaces carry two inequivalent irreducible representations
of ${\cal O}$, called $F_1,F_2$, see \cite{LF}. According to the considerations
in section 5 we may conjecture that these two
representations are generated by stationary,
symmetric, non-coplanar states.\\

Indeed, this can be directly verified for the following states:
Let $\bi{r}_\mu, \mu=1,\ldots,8$ denote the unit vectors pointing to the vertices
of the cube. The the states
\begin{equation}\label{28}
\bi{s}_\mu^{(1)}\equiv \bi{r}_\mu,\quad \mu=1,\ldots,8,
\end{equation}
and
\begin{equation}\label{29}
\bi{s}_\mu^{(-1)}\equiv (-1)^\mu \bi{r}_\mu,\quad \mu=1,\ldots,8,
\end{equation}
have the desired properties. They are easily vizualized:
$\bi{s}^{(1)}$ is just the ``cube in spin space" and
$\bi{s}^{(-1)}$ the tetrahedron where each of the four distinct spin vectors
is attached to pairs of vertices of the cube connected by space diagonals.
$\bi{s}^{(1)}$ is the anti-ground state for $S=0$ with energy $E_{max}(0)=8J$.
Since it is not coplanar, theorem 1 is not directly applicable.
However, it is possible to project the cube $\bi{s}^{(1)}$ onto a square
and, in a second step, to enlarge the square to a square of unit vectors.
Since these are linear transformations, the resulting co--planar state
$\bi{s}^{(1)\prime}$  is weakly symmetric and
has the same energy as before, namely $8J=E_{max}(0)$.
Now theorem 1 yields $E_{\small max}(S)=J(\frac{1}{4}S^2+8)$.

\subsubsection{The octahedron}

As noted in \cite{AL}, the octahedron can be decomposed into four triangles
with disjoint edges and it is 3-colorable. Hence it has coplanar ground states
with $S=0$
which are obtained by extensions of the local ground states of the triangles.
Since $j=4 J$ and $j_{{\small min}}=-2 J$ we again obtain the RBS-parabola
\begin{equation}\label{30}
E_{{\small min}}(S)=J(S^2-12).
\end{equation}
The eigenspace of ${\Bbb J}$ corresponding to $j_{{\small min}}$ is two-dimensional and
carries a real, irreducible representation of ${\cal O}$. This corresponds to the
one-dimensional complex eigenspace of ${\Bbb J}$  spanned by the vector
\begin{equation}\label{31}
\bi{s}_\mu=z_{n(\mu)},\quad \mu=1,\ldots,6,
\end{equation}
where the
\begin{equation}\label{32}
z_n\equiv\exp(i n 2 \pi/3),\quad n=0,1,2,
\end{equation}
are the $3$rd roots of unity and $\mu\mapsto n(\mu)$ denotes any $3$-coloring
of the octahedron. It follows that the state $\bi{s}$ is symmetrical. \\

The remaining $3$-dimensional eigenspace of ${\Bbb J}$ with eigenvalue $j_{\small maxi}=0$
carries the $3$-dimensional
self-representation of ${\cal O}$ and  hence corresponds to the symmetric anti-ground
state with $S=0$ which can be vizualized as a octahedron in spin space.
Its energy is $E_{\small max}=0$.
Again, as in the case of the cube, there is a weakly symmetric co--planar
state with the same energy and theorem 1 yields
$E_{\small max}(S)=\frac{2J}{3}S^2$. This state has the form of a
(not necessarily regular) hexagon
such that opposing vertices of the octahedron in real space are mapped
onto opposing vertices of the hexagon in spin space.\\

The above RBS energy bounds can also be obtained more simply:
Since $H_0={\bi S}^2-({\bi S}_{16}^2+{\bi S}_{25}^2+{\bi S}_{34}^2)$,
where ${\bi S}_{ij}\equiv{\bi S}_i+{\bi S}_j$ and $J=1$, the minimal energy
$E_{\small min}(S)$ is obtained for $S_{ij}=2$ as
$E_{\small min}(S)= S^2-12$. Similarly the energy is maximal for
$S_{ij}=\frac{1}{3}S$ which yields
$E_{\small max}(S)= S^2(1-\frac{3}{9})=\frac{2}{3}S^2$.

\subsubsection{The cuboctahedron}

\begin{figure}[f05]
\begin{center}
\epsfig{file=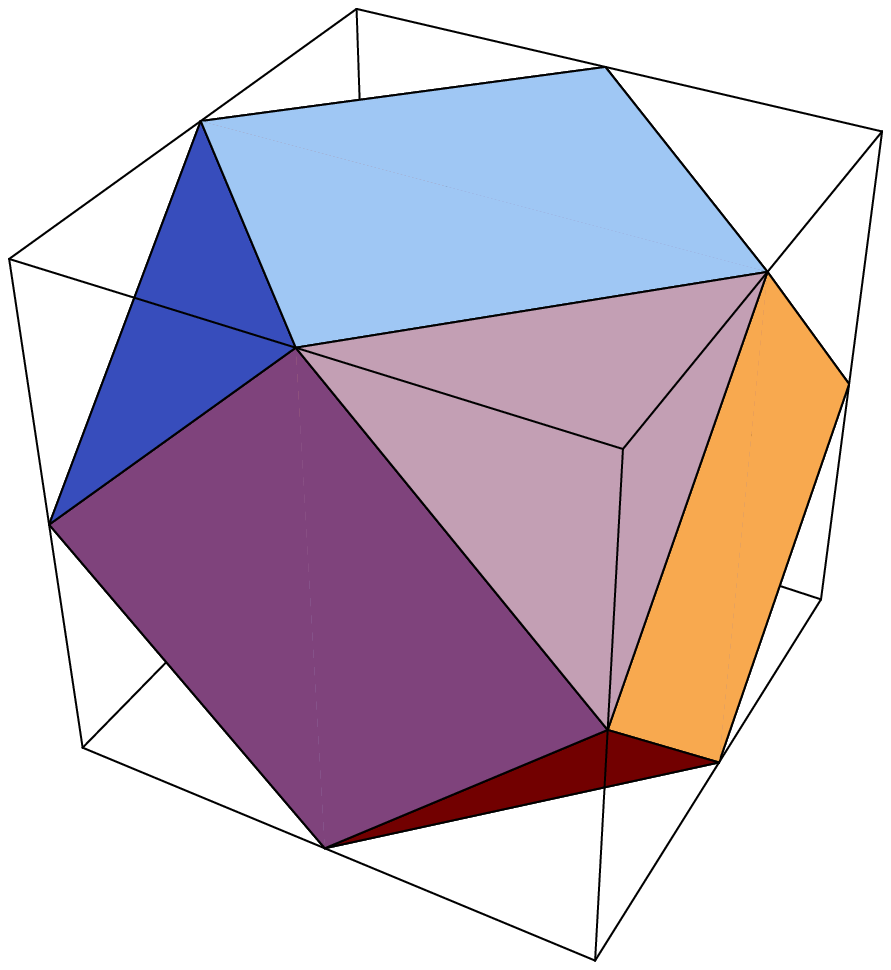,width=100mm}
\vspace*{1mm}
\caption[]{The cuboctahedron (shaded figure) is obtained by joining the mid-points
of the cube's edges with their nearest neighbors.}
\label{F-5}
\end{center}
\end{figure}

The cuboctahedron is the quasi-regular polyhedron obtained by joining the mid-points
of the cube's edges with their nearest neighbors, see \cite{Cox}, \cite{MW}, and figure \ref{F-5}.

It can be
decomposed into $8$ triangles with disjoint edges and it is 3-colorable, see figure 2.
Hence there is a coplanar ground state with $S=0$ giving rise to an RBS-parabola
\begin{equation}\label{33}
E_{{\small min}}(S)=J(\frac{1}{2}S^2-24),
\end{equation}
since $j=4$ and $j_{{\small min}}=-2$. The eigenvalues of ${\Bbb J }$ together with their
degeneracy and irreducible representations of ${\cal O}$ are summarized in table 1.

\begin{table}\label{Table1}
\caption{Eigenvalues $j_n$ of the adjacency matrix ${\Bbb J}$ of the cuboctahedron
(1st column) together with their degeneracy (2nd column). In the 3rd column
the irreducible representations of ${\cal O}$ are indicated which occur in the corresponding eigenspaces.
The nomenclature $A_1,A_2,E,F_1,F_2$ follows \cite{LF}.
}
\begin{indented}
\item[]\begin{tabular}{@{}rll}
\br
$j_n/J$ & degeneracy  &  irreducible representations of  ${\cal O}$ \\
\mr
$-2$ & $5$ & $E\oplus F_2$\\
$0$ & $3$ & $F_2$\\
$2$ & $3$ & $ F_1$\\
$4$ & $1$ & $A_1$\\
\br
\end{tabular}
\end{indented}
\end{table}

This has been calculated
by using the formula connecting the characters with the multiplicity of irreducible representations,
c.~f.~\cite{LF}, 4.2.31b. From this table it is obvious that the trivial irreducible representation $A_1$
corresponds to the total anti-ground state, $E$ is spanned by the
complex eigenvector of the coplanar ground state, and the self-representation
$F_1$ corresponds to the cuboctahedron in spin space with the energy
$E_{max}(0)=12 j_{\small maxi} = 24 J$. It remains to identify the symmetrical states
which generate the two 3-dimensional irreducible representations $F_2$ corresponding to the eigenvalues
$0$ and $-2$ of ${\Bbb J}$.\\

\begin{figure}[f08]
\begin{center}
\epsfig{file=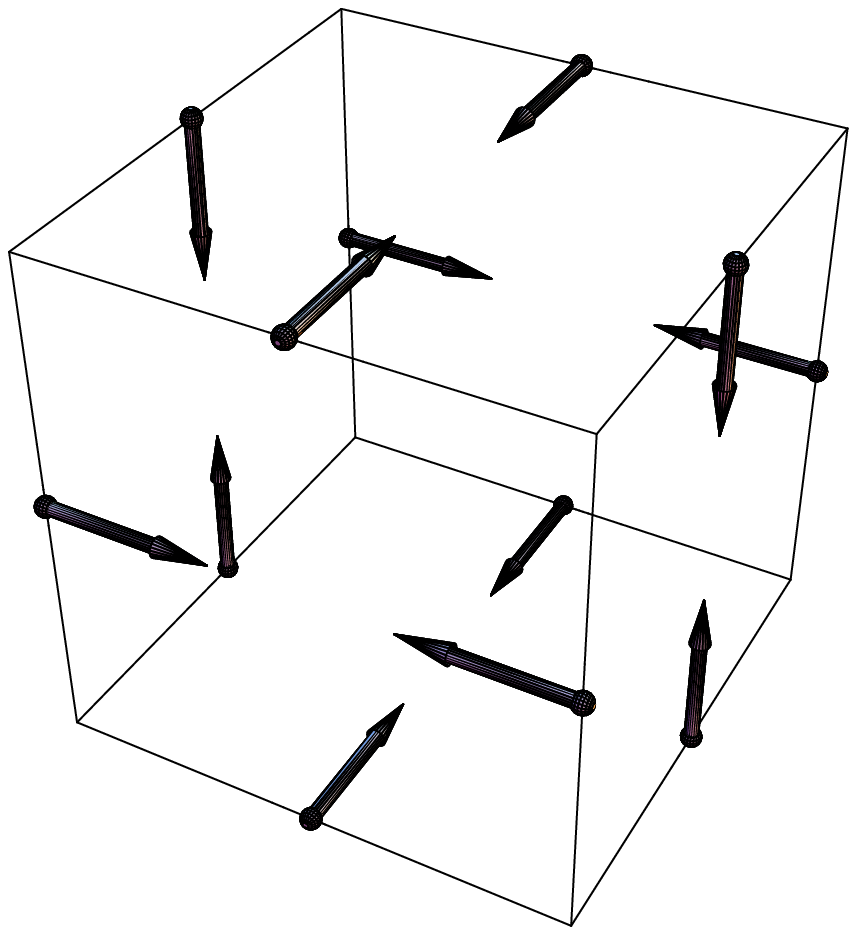,width=100mm}
\vspace*{1mm}
\caption[]{A stationary state of the cuboctahedron with energy $E=0$. The spin vectors
${\bi s}_\mu$ are shown as small arrows attached to vertices of the cuboctahedron.}
\label{F-8}
\end{center}
\end{figure}

The vertex vectors $\bi{r}_\mu, \mu=1,\ldots,12$ of the cuboctahedron
may be represented, up to normalization, by integer $3$-vectors
with components $-1,0,1$ and exactly one $0$-component. Then a state vector $\bi{s}_\mu$
may be defined by the following rule: Invert the first component of $\bi{r}_\mu$
after $0$ and set the other two components to $0$. Here ``after" is understood
cyclically, e.~g.~$(1,1,0)\mapsto (-1,0,0)$. Thus we obtain a state, see figure \ref{F-8}
where the spins $\bi{s}_\mu$ of adjacent vertices are orthogonal. Indeed, this state
corresponds to the $3$-dimensional eigenstate of ${\Bbb J}$ with eigenvalue $0$
which is transformed under ${\cal O}$ according to $F_2$.\\

A second state $\bi{s'}$ is obtained by a permutation of the $\bi{r}_\mu$. Each
$\bi{r}_\mu$ is mapped onto that vertex where the above-defined state vector
$\bi{s}_\mu$ points to:
\begin{equation}\label{34}
\bi{s'}_\mu\equiv 2\bi{s}_\mu -\bi{r}_\mu, \quad \mbox{(not normalized)}
\end{equation}
Alternatively, $\bi{s'}_\mu$ is obtained from $\bi{r}_\mu$ by the following rule:
Invert the first component of $\bi{r}_\mu$ after $0$ and leave the other unchanged.
For example, $(1,1,0)\mapsto(-1,1,0)$. Also this state generates a $3$-dimensional
subspace of the eigenspace of ${\Bbb J}$ with $j_{{\small min}}=-2J$ which transforms
under ${\cal O}$ according to $F_2$.

\begin{figure}[f06]
\begin{center}
\epsfig{file=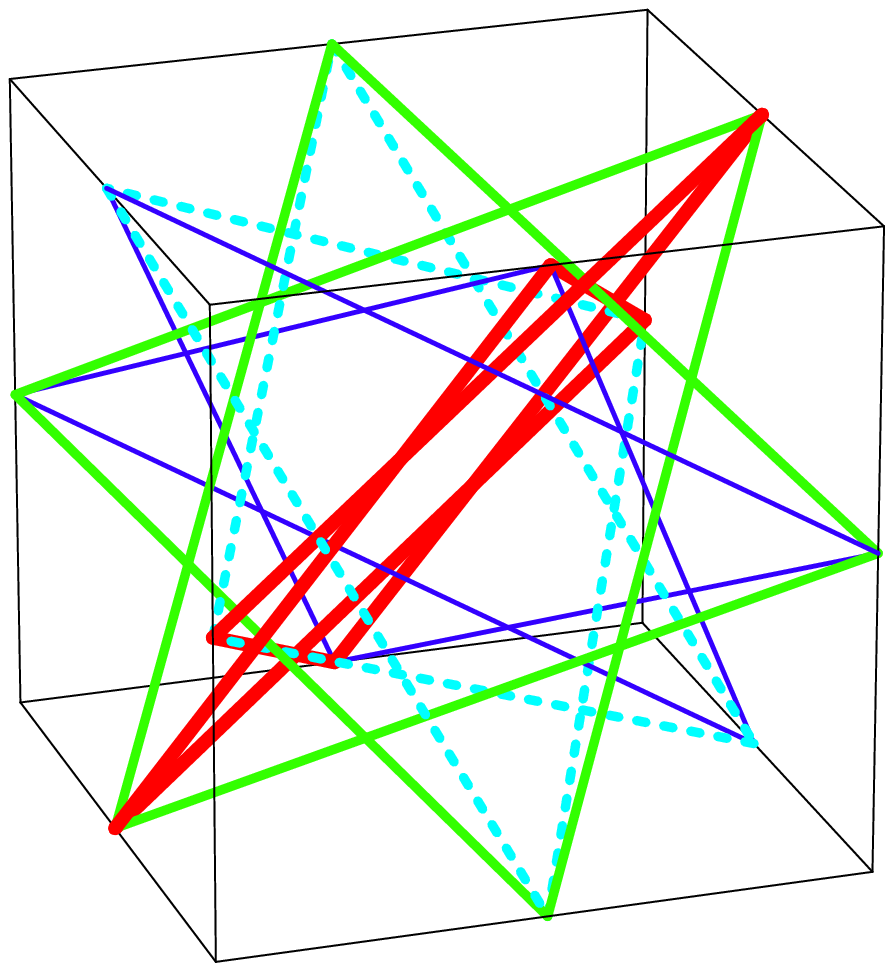,width=100mm}
\vspace*{1mm}
\caption[]{A ground state of the cuboctahedron formed by $4$ Stars of David.}
\label{F-6}
\end{center}
\end{figure}

Geometrically, $\bi{s'}$ is a figure in spin space formed by four Stars of David,
see figure \ref{F-6}. Thus the cuboctahedron has a coplanar ground state with $S=0$ as well as a non-coplanar
one with the same energy.\\

\subsection{${\cal Y}$-symmetrical systems}
We will consider the two remaining Platonic solids as well as the quasi-regular
icosidodecahedron and again calculate the decomposition of the eigenspaces of
the adjacency matrix ${\Bbb J}$ into ${\cal Y}$-irreducible subspaces for these three
cases. As in the case of ${\cal O}$-symmetric systems, the eigenvalue $j_{\small maxi}$
always corresponds to the $3$-dimensional self-representation of ${\cal Y}$ called $F_1$.
However, we have not found coplanar states realizing $j_{\small maxi}$ and hence
the question whether $E_{\small max}(S)$ will be a parabola remains open.\\

The situation for ground states is different from the previous examples: For
the icosahedron and the dodecahedron we find symmetric ground states with $S=0$ corresponding
to the $3$-dimensional irreducible representation $F_2$ of ${\cal Y}$,
but no coplanar ones. Moreover, numerical calculations yield the results
\begin{eqnarray}\label{33}
E_{\small min}^{\small coplanar}\approx -43.0614\ldots J
&>&
E_{\small min}=-20 \sqrt{5}J
\mbox{ for the dodecahedron,}\\
E_{\small min}^{\small coplanar}\approx -24 J
&>&
E_{\small min}=-12 \sqrt{5}J
\mbox{ for the icosahedron.}
\end{eqnarray}
Hence we
conjecture that there are no coplanar ground states and hence, according to theorem 2,
$E_{\small min}(S)$ will not be an exact parabola. Of course,
even overwhelming numerical evidence cannot be considered as a rigorous proof.
We can only strictly exclude symmetric coplanar
ground states, since there are no two-dimensional irreducible representations of ${\cal Y}$. This is
in accordance with proposition \ref{P2} and
the fact that the recently discovered $S=0$ ground states of the
icosidodecahedron \cite{AL} are coplanar and weakly symmetric, but not symmetric.\\

For the two ${\cal Y}$-symmetrical Platonic solids we will investigate more closely the
$3$-dimensional
geometry of the ground states. By numerical simulation of a heat bath at zero temperature,
Ch.~Schr\"oder \cite{CS} has found the angles between adjacent spins to be
$\alpha_I\approx 116.6^\circ$ for the icosahedron and
$\alpha_D\approx 138.2^\circ$ for the dodecahedron.
From the knowledge of the corresponding irreducible subrepresentation of the
natural representation of  ${\cal Y}$ in ${\Bbb R}^N$ we can now exactly
determine the symmetric ground state, see proposition \ref{P4}.
It turns out that these
states are well-known stellated geometrical structures, called ``Great Icosahedron" for
the ground state of the icosahedron and ``Great Stellated Dodecahedron" for the ground state
of the dodecahedron, see \cite{Cox} and \cite{MW}.  Thus the angles
between adjacent spins are just the angles
between neighboring vertices of these stellated structures and hence have the exact values
\begin{equation}
\alpha_I=\arccos(-\textstyle\frac{\sqrt{5}}{5}),\;
\alpha_D=\arccos(-\textstyle\frac{\sqrt{5}}{3}),\end{equation}
in agreement with the numerical findings of \cite{CS}.
\begin{table}\label{Table2}
\caption{Eigenvalues $j_n$ of the adjacency matrix ${\Bbb J}$ of the icosahedron
(1st column) together with their degeneracy (2nd column). In the 3rd column
the irreducible representations of ${\cal Y}$ are indicated which occur in the corresponding eigenspaces.
The nomenclature $A,F_1,F_2,G,H$ follows \cite{LF}.
}
\begin{indented}
\item[]\begin{tabular}{@{}rll}
\br
$j_n/J$ & degeneracy  &  irreducible representations of  ${\cal Y}$ \\
\mr
$-\sqrt{5}$ & $3$ & $F_2$\\
$-1$ & $5$ & $H$\\
$\sqrt{5}$ & $3$ & $F_1$\\
$5$ & $1$ & $A$\\
\br
\end{tabular}
\end{indented}
\end{table}
\begin{table}\label{Table3}
\caption{Eigenvalues $j_n$ of the adjacency matrix ${\Bbb J}$ of the dodecahedron
(1st column) together with their degeneracy (2nd column). In the 3rd column
the irreducible representations of ${\cal Y}$ are indicated which occur in the corresponding eigenspaces.
The nomenclature $A,F_1,F_2,G,H$ follows \cite{LF}.
}
\begin{indented}
\item[]\begin{tabular}{@{}rll}
\br
$j_n/J$ & degeneracy  &  irreducible representations of  ${\cal Y}$ \\
\mr
$-\sqrt{5}$ & $3$ & $F_2$\\
$-2$ & $4$ & $G$\\
$0$ & $4$ & $G$\\
$1$ & $5$ & $H$\\
$\sqrt{5}$ & $3$ & $F_1$\\
$3$ & $1$ & $A$\\
\br
\end{tabular}
\end{indented}
\end{table}
\begin{table}\label{Table4}
\caption{Eigenvalues $j_n$ of the adjacency matrix ${\Bbb J}$ of the icosidodecahedron
(1st column) together with their degeneracy (2nd column). In the 3rd column
the irreducible representations of ${\cal Y}$ are indicated which occur in the corresponding eigenspaces.
The nomenclature $A,F_1,F_2,G,H$ follows \cite{LF}.
}
\begin{indented}
\item[]\begin{tabular}{@{}rll}
\br
$j_n/J$ & degeneracy  &  irreducible representations of  ${\cal Y}$ \\
\mr
$-2$ & $10$ & $H\oplus H$\\
$1-\sqrt{5}$ & $3$ & $F_2$\\
$-1$ & $4$ & $G$\\
$1$ & $4$ & $G$\\
$2$ & $5$ & $H$\\
$1+\sqrt{5}$ & $3$ & $F_1$\\
$4$ & $1$ & $A$\\
\br
\end{tabular}
\end{indented}
\end{table}

\section{Summary}
We have tried to give a systematic survey of the properties of (relative) ground states
of classical Heisenberg spin systems with particular emphasis on symmetric systems.
Further we have devised various methods of ground state construction, e.~g.~extension of
local ground states, construction of symmetric ground states from irreducible
representations of the system's symmetry group, and the construction of relative
(anti-) ground states from coplanar ground states with $S=0$.
The latter procedure yields upper and lower parabolas as the boundaries of the
$S$-resolved energy spectrum. Thus we have a sufficient condition for
a system to be parabolic. Moreover, under certain assumptions, this condition can
also be proved to be necessary.\\

The above issues are also relevant for the quantum
theory of Heisenberg spin systems. It has been shown for various cases \cite{JM},\cite{JS}
that the shape of the $S$-resolved energy spectrum of the quantum system is very well
approximated by the curves $E_{\small min}(S)$ and $E_{\small max}(S)$ of the
corresponding classical system, if only the individual spin quantum number $s$ exceeds
some moderate value, say $s=2$.
Hence one can predict semi-quantitative features
of the quantum spectrum if one knows the classical spectrum and
$s \raisebox{-0.5ex}{$\stackrel{>}{\sim}$}2$.

For some quantum systems, like the icosidodecahedron with $s=5/2$, where the diagonalization of Heisenberg
Hamiltonian is totally impractical, approximation methods such as the density matrix renormalization group
method (DMRG) are currently able to provide estimates for the lower boundary of the eigenvalue spectrum. In
fact, for the icosidodecahedron the estimates \cite{ES} are in very good quantitative
agreement with the lower boundary of the parabolic spectrum of the classical system. This agreement suggests
that the DMRG results are also close to those for the quantum system with $s=5/2$.
For other classical systems with the same symmetry group ${\cal Y}$, viz.~the dodecahedron
and the icosahedron we expect non-parabolicity, although the corresponding quantum
systems may exhibit approximate rotational bands.\\

In this sense our results and case studies are to a great extent relevant also for real,
quantum spin systems.

\section*{Acknowledgements}
H.-J.~S.~would like to thank members of the Condensed Matter Physics Group of
the Ames Laboratory
for their warm hospitality during a visit when a part of
this work was performed.
Likewise, M.~L.~thanks members of Fachbereich Physik, Universit\"at Osnabr\"uck for their warm
hospitality when the final version of the manuscript was completed. We also
acknowledge the financial support of a travel grant awarded jointly by
NSF-DAAD which facilitated these visits.
Finally, it is a pleasure
to thank M.~Axenovich, K.~B\"arwinkel, J.~Schnack, and Ch.~Schr\"oder
for stimulating and helpful discussions. Ames Laboratory is operated for the United
States Department of Energy by Iowa State University under Contract
No. W-7405-Eng-82.

\end{document}